\documentclass[aps,prd,longbibliography,onecolumn,showpacs,amsmath,amssymb,superscriptaddress,nofootinbib,floatfix,preprintnumbers,secnumarabic]{revtex4-1}

\usepackage[usenames,dvipsnames]{color}
\usepackage{xcolor}
\usepackage{multirow}
\usepackage{graphicx}
\usepackage{adjustbox}
\usepackage{booktabs}
\usepackage{appendix}
\usepackage{commath}
\usepackage{mathtools}
\usepackage{enumitem}
\usepackage{slashed}
\usepackage{url}
\usepackage{xspace}
\usepackage{braket}
\usepackage{amsmath,amssymb,amsthm,amsfonts}
\usepackage{tabularx}
\definecolor{darkblue}{rgb}{0,0,0.5}
\definecolor{darkgreen}{rgb}{0.1,0,0.3}
\definecolor{darkred}{rgb}{0.6,0,0}
\usepackage[bookmarks=false, bookmarksnumbered=true, colorlinks=true, urlcolor=darkblue, citecolor=darkgreen, linkcolor=darkred, breaklinks=true]{hyperref} 
\usepackage[capitalise]{cleveref}
\usepackage[normalem]{ulem} 
\usepackage{siunitx}
\usepackage[utf8]{inputenc}
\usepackage{xspace}
\usepackage{bm}
\usepackage{braket}
\usepackage{tikz}
\usepackage{physics}

\usepackage{fullpage}
\usepackage{geometry}      
\AtBeginDocument{
  \heavyrulewidth=.08em
  \lightrulewidth=.05em
  \cmidrulewidth=.03em
  \belowrulesep=.65ex
  \belowbottomsep=0pt
  \aboverulesep=.4ex
  \abovetopsep=0pt
  \cmidrulesep=\doublerulesep
  \cmidrulekern=.5em
  \defaultaddspace= .5em
  \setlength{\tabcolsep}{0.5em}
}

\newcommand{\be}{\begin{equation}\begin{aligned}}
\newcommand{\ee}{\end{aligned}\end{equation}}
\newcommand{\beq}{\begin{equation}}
\newcommand{\eeq}{\end{equation}}
\newcommand{\beqa}{\begin{eqnarray}}
\newcommand{\eeqa}{\end{eqnarray}}

\newcommand{\OmegaDM}{\Omega_{\text{DM}}}

\newcommand{\gev}{\text{GeV}}
\newcommand{\tev}{\text{TeV}}

\newcommand{\Eqref}[1]{Equation~(\ref{#1})}

\renewcommand{\eqref}[1]{Eq.~(\ref{#1})}

\newcommand{\mpl}{\ensuremath{M_\text{P}}}

\newcommand{\ms}{\ensuremath{m_\sigma}}

\newcommand\Tstrut{\rule{0pt}{4ex}}   

\RequirePackage[normalem]{ulem}

\begin{document}

\title{Dark matter production through a non-thermal flavon portal}

\author{Andrew Cheek}
\email{acheek@camk.edu.pl}
\affiliation{Astrocent, Nicolaus Copernicus Astronomical Center Polish Academy of Sciences, ul.~Rektorska 4, 00-614, Warsaw, Poland}
\vspace*{0.25in}

\author{Jacek K. Osi\'nski}
\email{josin@camk.edu.pl}
\affiliation{Astrocent, Nicolaus Copernicus Astronomical Center Polish Academy of Sciences, ul.~Rektorska 4, 00-614, Warsaw, Poland}
\vspace*{0.25in}

\author{Leszek Roszkowski}
\email{leszek.roszkowski@ncbj.gov.pl}
\affiliation{Astrocent, Nicolaus Copernicus Astronomical Center Polish Academy of Sciences, ul.~Rektorska 4, 00-614, Warsaw, Poland}
\affiliation{National Centre for Nuclear Research, ul.~Pasteura 7, 02-093 Warsaw, Poland
\vspace*{0.25in}
}

\author{Sebastian Trojanowski
\vspace*{0.2in}
}
\email{Sebastian.Trojanowski@ncbj.gov.pl}
\affiliation{Astrocent, Nicolaus Copernicus Astronomical Center Polish Academy of Sciences, ul.~Rektorska 4, 00-614, Warsaw, Poland}
\affiliation{National Centre for Nuclear Research, ul.~Pasteura 7, 02-093 Warsaw, Poland
\vspace*{0.25in}
}


\begin{abstract}
The Froggatt-Nielsen (FN) mechanism provides an attractive way of generating the determined fermion mass hierarchy and quark mixing matrix elements in the Standard Model (SM). Here we extend it by coupling the FN field, the flavon, to a dark sector containing one or more dark matter particles which are produced non-thermally sequentially through flavon production. Non-thermal flavon production occurs efficiently via freeze-in and through field oscillations. We explore this in the regime of high-scale breaking $\Lambda$ of the global $U(1)_{\textrm{FN}}$ group and at the reheating temperature $T_R\ll \Lambda$ where the flavon remains out of equilibrium at all times. We identify phenomenologically acceptable regions of $T_R$ and the flavon mass where the relic abundance of dark matter and other cosmological constraints are satisfied. In the case of one-component dark matter we find an effective upper limit on the FN charges at high $\Lambda$, i.e. $Q_{\rm FN}^{\rm DM}\leq13$. In the multi-component dark sector scenario the dark matter can be the heaviest dark particle that can be effectively stable at cosmological timescales, alternatively it can be produced sequentially by decays of the heavier ones. For scenarios where dark decays occur at intermediate timescales, i.e. $t\sim 0.1- 10^{28}\,{\rm s}$, we find that existing searches can effectively probe interesting regions of parameter space. These searches include indirect probes on decays such as $\gamma$-ray and neutrino telescopes as well as analyses of the Cosmic Microwave Background, as well as constraints on small scale structure formation from the Lyman-$\alpha$ forest. We comment on the future prospects of such probes, place projected sensitivities, and discuss how this scenario could accommodate the cosmological $S_8$ tension. 
\end{abstract}


\maketitle

\section{Introduction}
\label{sec:introduction}

The success of the Standard Model (SM) of particle physics in explaining fundamental interactions has been well established over the last several decades with a remarkable number of experimental observations. Still, however, our understanding of nature at its most fundamental level suffers from the lack of answers to many important questions. Among them is the perplexing hierarchy in the Higgs boson interaction strength with various SM fermions which is encoded in the Yukawa couplings, known as the flavor puzzle, as well as observed phenomena that strongly motivate searches for physics beyond the SM (BSM). The most striking example of the latter is indicated by the cumulative evidence for the existence of dark matter (DM) which contributes to the energy budget of the Universe in a way exceeding that of ordinary baryonic matter.

The flavor puzzle has also been addressed with various possible extensions of the SM, cf. Refs~\cite{Feruglio:2015jfa,Isidori:2015oea,Altmannshofer:2022aml} for reviews. A particularly popular attempt utilizes the observation that the hierarchy in the masses and mixing matrix elements among the SM fermions can be conveniently reproduced up to order-one factors with simple integer powers of a common quantity $\epsilon\simeq 0.23$, which is of the order of the Cabibbo angle. This gave rise to the well-known framework of Froggatt-Nielsen (FN) models~\cite{Froggatt:1978nt}, cf. also Refs~\cite{Wilczek:1982rv,Berezhiani:1989fp,Berezhiani:1990wn,Berezhiani:1990jj,Leurer:1992wg,Leurer:1993gy,Ibanez:1994ig,Sakharov:1994pr,Sato:1997hv,Irges:1998ax,Fedele:2020fvh}. The simplest ultraviolet (UV) completions of the FN scenario postulate the existence of additional vector-like fermions with mass of order $\Lambda$ which is typically required to lie well above the $\tev$ scale to avoid otherwise stringent bounds from flavor-changing-neutral-current (FCNC) processes. The new fermions are coupled non-universally to the SM fermions as dictated by associated charges with respect to an Abelian flavor group $U(1)_{\textrm{FN}}$. At lower energies, once the heavy fields are integrated out, the Yukawa coupling can be described in terms of powers of $\epsilon\sim \langle S\rangle/\Lambda$, where $\langle S\rangle$ is the vacuum expectation value (vev) of an additional SM-singlet scalar field $S$, dubbed the flavon. In more complex cases, a number of flavon fields charged under $U(1)_{\textrm{FN}}$ can exist and develop a non-zero vev after the flavor symmetry breaking. In particular, various realizations of FN-type scenarios consistent with the Grand Unified Theories (GUTs) have been proposed to accommodate a range of SM data from heavy quarks to neutrino mixings~\cite{Altarelli:2002sg,Buchmuller:2011tm,Altarelli:2012ia,Ding:2012wh}. 

As far as the nature of DM is concerned, a plethora of theoretical scenarios have been analyzed that predict the existence of DM particles with mass spanning many orders of magnitude between $m_{\textrm{DM}}\sim 10^{-21}~\textrm{eV}$, characteristic for fuzzy DM, to macroscopic primordial black holes. In particular, many leading DM models employ a popular thermal production mechanism and postulate the existence of new weakly-interacting massive particles (WIMPs) with mass of order the weak scale, cf. Refs.~\cite{Arcadi:2017kky,Roszkowski:2017nbc,Cooley:2022ufh} for recent reviews. The precise abundance of such DM particles can then be predicted in the standard cosmological scenario by fixing their interaction strength with the SM species. On the other hand, the lack of observation of DM interactions in direct and indirect searches, cf. Refs~\cite{Billard:2021uyg,Boveia:2022syt}, puts growing pressure on the thermal DM paradigm and motivates further research on non-thermal production mechanisms. These could explain the observed DM abundance while avoiding the aforementioned bounds due to much-suppressed couplings between the dark and SM species, for reviews see Refs.~\cite{Baer:2014eja,Bernal:2017kxu}.

While it is commonly assumed that the current DM relic abundance corresponds to the presence of a single stable new species $\chi$, this particle may arise from a more complicated dark sector. This is also taken into account when estimating the $\chi$ relic density as heavier dark species might decay into $\chi$ or otherwise contribute to its production rates in the early Universe. Such decays may also happen at much later times and lead to semi-visible signatures or even very complicated evolution of the relic densities of dark sector species, cf., e.g. a dynamical DM scenario~\cite{Dienes:2011ja,Dienes:2011sa}. The phenomenology of such models is then largely dictated by the mass hierarchy in the dark sector and the relevant decay widths.

In this study, we explore such a BSM scenario with the hierarchy between the dark-sector particles driven by the vev of the flavon field $S$ which simultaneously generates the Yukawa couplings in the SM. It has been previously noticed that such a flavor-portal model can lead to thermal DM production in the early Universe with its freeze-out driven by annihilations through an intermediate flavon, $\bar{\chi}\chi\to S^\ast\to f\bar{f}$, where $f$ are the SM fermions~\cite{Calibbi:2015sfa}, see also Refs.~\cite{Hirsch:2010ru,Boucenna:2011tj,Kile:2011mn,Batell:2011tc,Kamenik:2011nb,Agrawal:2011ze,Lopez-Honorez:2013wla,Batell:2013zwa,Agrawal:2014una,Agrawal:2014aoa,Bishara:2015mha,Agrawal:2015kje,deMedeirosVarzielas:2015lmh,Bhattacharya:2016lts,Galon:2016bka,Bhattacharya:2016rqj,Alvarado:2017bax,Baek:2017ykw,Bernal:2017xat,Renner:2018fhh,Desai:2020rwz} for further discussion about DM and flavor symmetries. Instead, for increasing values of the UV-completion scale $\Lambda\gg\tev$, the thermal DM production cross section becomes suppressed and the predicted relic density grows above the determined value until interaction rates become small enough for the DM particles to never achieve equilibrium with the SM species.

Our focus below is precisely on such theoretically well-motivated high-scale FN-type models in which $\Lambda$ can grow even up to the GUT scale. We show that, in this case, a minimal coupling of the flavon field to a new postulated dark fermion in a way inspired by the interactions in the SM sector can give rise to a good DM candidate which avoids current bounds and predicts correct relic abundance with a non-thermal origin. We identify the dark FN charges for which this mechanism can be realized either due to the initial freeze-in production of heavy flavons or due to flavon production dominated by oscillations induced by thermal corrections to the potential~\cite{Lillard:2018zts}, cf. also Ref.~\cite{Chen:2019wnk} for possible connections to baryogenesis. We then focus on more general dark sectors in which the mass hierarchy between multiple dark species is driven by different FN charges. We illustrate that the existence of non-minimal dark sectors coupled to the SM this way could render these scenarios observable via late-time decays of the heavier dark fermions, therefore providing a new window to indirectly study high-scale FN-type models. Further phenomenological implications of such decays throughout the cosmological evolution of the Universe are also discussed.

The paper is organized as follows. In \cref{sec:model} we briefly present the model under study. The flavon and DM production mechanisms are discussed in \cref{sec:production}. \Cref{sec:results} is devoted to results of our analysis for both the single and multi-component dark sectors and we conclude in \cref{sec:conclusions}. In \cref{app:massbasis} we detail the mass and coupling corrections coming from the rotation between the flavor and mass basis. We also present a benchmark three-component dark sector that could address the cosmological $S_8$ tension. \Cref{app:darkcharges} contains remarks about general dark charge assignments.

\section{Model\label{sec:model}}

\subsection{Froggatt-Nielsen framework}

At energy scales much below $\Lambda$ the effective flavon interaction terms with the SM fields are given by
\begin{equation}
\mathcal{L} = \sum_{ij}{y^u_{ij}\,\left(\frac{S}{\Lambda}\right)^{n_{ij}^u}}\,\bar{Q}_i\,\tilde{\Phi}\,u_j + \sum_{ij}{y^d_{ij}\,\left(\frac{S}{\Lambda}\right)^{n_{ij}^d}}\,\bar{Q}_i\,\Phi\,d_j,
\label{eq:L}
\end{equation}
where $\Phi$, $Q$, $u$ and $d$ are the usual Higgs, left-handed quark doublet and the up-type and down-type right-handed quark singlet fields respectively. The subscript runs over the three of generations of the SM, $y^{u/d}_{ij}$ are dimensionless couplings of order $1$, while integer powers $n_{ij}^{u/d}$ appear after integrating out heavy vector-like fermions and are dictated by combinations of FN charges of the SM fields. In the following, we will focus on the quark sector of the SM as it plays the dominant role for the flavon masses of our interest. At high energies, the Lagrangian in \cref{eq:L} should be modified by explicitly taking into account other heavy fields. As will be discussed below, however, our focus is on non-thermal production of flavons and scenarios for which the reheating temperature of the Universe is suppressed with respect to the UV-completion scale, $T_R\ll \Lambda$. The effective description in \cref{eq:L} is then sufficient for our purposes and it allows us to derive results largely independent of the details of such a completion.

After breaking the $U(1)_{\rm FN}$ group, the flavon field gets a non-zero vev $v_s$,
\begin{equation}
S = \frac{1}{\sqrt{2}}(v_s + \sigma + i\rho).
\label{eq:S}
\end{equation}
The hierarchy of the Yukawa couplings of the SM is then given by the powers of $\epsilon=v_s/(\sqrt{2}\Lambda)\simeq 0.23$. Flavon couplings to quarks and the Higgs boson before the electroweak phase transition (EWPT) can be obtained from \eqref{eq:L} after expanding around the $S$ vev
\begin{equation}
\mathcal{L}_{\rm before\atop EWPT}\supset\sum_{ij}{g^u_{ij}\sigma \bar{Q}_i\tilde{\Phi}u_j} + \sum_{ij}{g^d_{ij}\sigma \bar{Q}_i\Phi d_j},
\label{eq:Lbefore}
\end{equation}
where
\begin{equation}
g_{ij}^u \simeq y_{ij}^u\,\frac{n^u_{ij}}{v_s}\,\epsilon^{n^u_{ij}},\hspace{1cm}g_{ij}^d \simeq y_{ij}^d\,\frac{n^d_{ij}}{v_s}\,\epsilon^{n^d_{ij}}.
\label{eq:gbefore}
\end{equation}
After the EWPT, the non-renormalizable operators become suppressed with respect to renormalizable terms that appear once the Higgs field gets its vev, $v_\phi\simeq 246~\gev$. The flavon-quark-quark couplings are obtained after rotation to the mass basis $u_{L,R}$ and $d_{L,R}$ 
\begin{equation}
\mathcal{L}_{\rm after\atop EWPT}\supset \frac{1}{\sqrt{2}}\frac{v_\phi}{v_s}\sigma\left(\bar{u}_L\,U_{u}^\dagger\,(n^u f^u)W_{u}\,u_R + \bar{d}_L\,U_{d}^\dagger\,(n^d f^d)W_{d}\,d_R\right),
\label{eq:Lafter}
\end{equation}
where $f^{u,d}$ are the Yukawa matrices, $(n f) = n_{ij}f_{ij}$, and the biunitary diagonalization of the Yukawa matrices is performed by $f^{u,d}_{\rm diag} = U_{u,d}^\dagger\,f^{u,d}\,W_{u,d}$. We define a new set of parameters, $\tilde{g}_{ij}^{u/d}$ such that
\begin{equation}
\mathcal{L}_{\rm after\atop EWPT}\supset \tilde{g}_{ij}^{u}\sigma\bar{u}_L\,u_R +\tilde{g}_{ij}^{d}\sigma\bar{d}_L\,d_R.
\label{eq:Lafter_gtilde}
\end{equation}

The precise mass of the flavon field is determined by the parameters in its potential and would naturally be of order $m_\sigma \sim v_s\sim\Lambda$. In the case of a global $U(1)_{\textrm{FN}}$ symmetry, the pseudoscalar field $\rho$ in \cref{eq:S} would be a Nambu-Goldstone boson generating a long-range force between quarks. Alternatively, one could consider the $U(1)_{\rm FN}$ as a local symmetry, whereby $\rho$ will be `swallowed' by the additional gauge boson, much like in the Higgs mechanism. This, too, could introduce certain complications to the phenomenology, for example, often charge configurations are anomalous~\cite{Chankowski:2005qp}, where the usual workaround is that anomaly cancellation would occur with additional field content near $\Lambda$. The phenomenology of such a model will, however, be largely driven by this new vector boson whose couplings to the SM fields are not suppressed by $\Lambda$, cf. discussion in Ref.~\cite{Calibbi:2015sfa}. In order to lift the $\rho$ mass, the $U(1)_{\textrm{FN}}$ group could also be broken either explicitly~\cite{Baldes:2016gaf}, or by the QCD anomaly giving rise to the axiflavon solution to the strong CP problem~\cite{Calibbi:2016hwq,Ema:2016ops,Panci:2022wlc}, interestingly here the axion itself is a viable dark matter candidate. Alternatively, the $U(1)_{\rm{FN}}$ group could be replaced by a discrete $\mathbb{Z}_N$ symmetry which maintains the FN mechanism while allowing $m_\rho>m_\sigma$~\cite{Lillard:2018zts}. In the following, we will assume for concreteness that, indeed, the pseudoscalar mass is larger than the scalar one such that the phenomenology of the model is determined by flavon interactions. We note, however, that even if $m_\rho\lesssim \ms$, much of the discussion below could remain valid with the pseudoscalar $\rho$ playing a role similar to the flavon. Since we do not specify the precise flavon potential, in the following we will treat the flavon mass $m_\sigma$ as a free parameter which can also be driven much below $\Lambda$ thanks to tuning of the relevant parameters. Examples of pseudo-anomalous $U(1)_{\textrm{FN}}$ groups in which the masses of the scalar fields are typically much lower than the UV-completion scale of order the Planck mass $\mpl$ can also arise in the string-theory context~\cite{Binetruy:1994ru,Binetruy:1996xk}.

\subsection{Flavon coupling to dark matter}

Apart from heavy vector-like fermions in the UV-complete FN scenario, we assume that the dark sector is also populated with lighter fermions $\chi_i$, out of which the lightest one can be made stable and therefore can be a good DM candidate. The heavier dark fermions can decay into the lighter dark and visible states, as is typical in many BSM frameworks. 
 
If not forbidden by any additional symmetry in the dark sector, the simplest interaction term between these dark fermions and the flavon field is given by $S \bar{\chi}\chi$ which would naturally yield $m_\chi\sim v_s$. In the following, we assume that in more complicated dark sectors the mass hierarchy between the dark fermions is set via a similar mechanism to the SM sector. In the limit where heavy fields with mass of order $\Lambda$ can be integrated out, the relevant couplings are then augmented with additional $(S/\Lambda)$ terms
\begin{equation}
\mathcal{L} \supset \sum_{ij}{y^\chi_{ij}\,\left(\frac{S}{\Lambda}\right)^{n_{ij}^\chi}}\,S\,\bar{\chi}_i\,\chi_j,
\label{eq:Lchi1}
\end{equation}
where, for concreteness, we set $y^\chi_{ii} = 1$ and $y^\chi_{ij} \equiv y = 1/3$ for off-diagonal coefficients with $i\neq j$. We stress that our results are only slightly affected by changing the $\mathcal{O}(1)$ coefficient $y$.\footnote{The notable exception can be obtained by setting $y = 1$, for which the determinant of the mass matrix obtained by expanding \cref{eq:Lchi1} around the vev of $S$ vanishes and the lighter dark species becomes dark radiation.} Here, for $n_{ij}^\chi = 0$ we reproduce the direct interaction term between $S$ and $\chi_i$, while for larger values of $n_{ij}^\chi$ the corresponding masses and coupling strengths become suppressed by powers of $\epsilon$. While we keep the number of dark fermions $\chi_i$ unspecified, below we illustrate that important aspects of the phenomenology of such scenarios can be well captured by discussing one or two such fields with specific dark FN charges.

We note the presence of an additional flavon field in \cref{eq:Lchi1} relative to \cref{eq:L}. While such terms are not allowed in the SM, here they provide the leading contribution to the dark fermion masses. These mass parameters are then set by $v_s\gg v_h$ and powers of $\epsilon$, and they are naturally larger than the masses of the SM species obtained for similar values of the FN charges
\begin{equation}
    m^{\chi}_{ij}=\frac{y_{ij}^\chi v_{s}}{\sqrt{2}}\left(\frac{v_{s}}{\sqrt{2}\Lambda}\right)^{n^{\chi}_{ij}} = y_{ij}^\chi\,\epsilon^{{n^{\chi}_{ij}+1}}\Lambda.
    \label{eq:mdm_eq}
\end{equation}
Similarly to the SM, the dark fermions with lower values of the charges are heavier and they couple more strongly to the flavon field
\begin{equation}
    \mathcal{L}\supset \sum_{ij}\frac{y_{ij}^\chi\epsilon^{n^{\chi}_{ij}}}{\sqrt{2}}\left(n^{\chi}_{ij}+1\right)\,\sigma\bar{\chi}_i\chi_j=\sum_{ij}g^\chi_{ij}\sigma\bar{\chi}_i\chi_j, 
    \label{eq:flavon_chichi_int}
\end{equation}
where we introduce the coupling $g^{\chi}_{ij}$ for ease of use later. Upon rotation to the mass basis, we obtain 
\begin{equation}
\mathcal{L} \supset \frac{1}{\sqrt{2}\,\epsilon\,\Lambda}\,\sigma
\,(\bar{\chi}\,R)\,(n^{\chi}\,m)(R^{-1}\,\chi),
\end{equation}
where $\bar{\chi} = (\chi_1\ \chi_2)$, $(n^\chi\,m)_{ij} = n^\chi_{ij}\,m_{ij}$, and $R$ is the relevant rotation matrix. We note that the ``$+1$'' term in \cref{eq:flavon_chichi_int}, which is not proportional to $n^\chi_{ij}$, vanishes in the mass basis. The mixing between the dark flavor states introduces additional corrections to the couplings $g_{ij}^\chi$ to the flavon field defined above. These are, however, not substantial for at least unit charge difference between the dark species, cf. \cref{app:massbasis} for further discussion.

The charge assignments in the chiral basis are given by 
\begin{equation}
    n^{\chi}_{ij}= \mathcal{Q}_{\chi_{i, L}} + \mathcal{Q}_{\chi_{j, R}} -1.
\label{eq:chargeschi}
\end{equation}
In this work, we assume $\mathcal{Q}_{\chi_{i, L}}=\mathcal{Q}_{\chi_{i, R}}$ for simplicity, cf. \cref{app:darkcharges} for remarks about going beyond this assumption. Additionally, as is the norm in the FN framework, we only choose integer charge assignments. For a multi-component dark sector, populated only by vector-like fermions, this yields a mass spectrum in which for low values of $n^{\chi}_{ij}$, the heaviest dark states will have masses just below the cut-off scale $\Lambda$. The masses of the lighter fermions $\chi_i$ can span many orders of magnitude below and are characterized by a characteristic minimum mass gap between neighbor fermions, $m_{\chi_{i+1}}\approx 19 \, m_{\chi_i}$. This mass difference can be lower, however, in case two or more dark fermions carry the same FN charge.

\section{Flavon and dark matter production\label{sec:production}}
\subsection{Flavons through oscillations and freeze-in}
\label{subsec:flavon_prod}

As mentioned above, for large values of the UV-completion scale $\Lambda$, and as long as the reheating temperature remains much lower, $T_R\ll \Lambda$, the effective description of flavon interactions in \cref{eq:L,eq:Lchi1} is valid for the conditions in the early Universe. In this case, the  flavon yield essentially depends on $\Lambda$ and $T_R$, as well as on the flavon mass $m_\sigma$, while further details of the complicated full BSM scenario can be encoded in the FN charges. These play a less important role given that the freedom in choosing FN textures is largely limited by the requirement of fitting the SM masses and entries of the CKM matrix. In this attractive and predictive regime of the model, flavon production relies on two main mechanisms, namely freeze-in  from the SM species in the thermal plasma and flavon oscillations driven by thermal corrections to its potential, cf. Ref.~\cite{Lillard:2018zts} for a detailed discussion, Refs~\cite{Buchmuller:2003is,Buchmuller:2004xr} for analogous remarks about the dilaton potential, and Ref.~\cite{Batell:2021ofv,Batell:2022qvr} for recent related studies about DM production. Below, we briefly recap the corresponding results.\\

We first note that flavons will be produced from the hot SM bath present in the early Universe via freeze-in. Depending on the temperature at a given time, there are two phases of production, corresponding to periods before and after the EWPT. Before the EWPT, we see from \Eqref{eq:Lbefore} that the flavon interacts with the Standard Model via a non-renormalisable (dimension 5) 4-point vertex. The production channels, therefore, come by way of $2\rightarrow 2$ scatterings. The flavon yield is given by
\begin{equation}
    Y^{u/d, {\rm UV}}_{ij} = \frac{3 \left\vert g_{ij}^{u/d}\right\vert^2 A_{\star} \mpl \ms}{64\pi^5}\int_{x_{\rm min}}^{x_{\rm max}}{\rm d}x\left(16-3x^2\right)K_{2}(x) + 8x K_1(x) + 3x^2 K_0(x),
    \label{eq:flavY_bwept}
\end{equation}
where $A_{\star}\equiv 45/(2\pi^2g_\star^S 0.33\sqrt{g_{\star}^{\rho}})$ is defined for convenience, $\mpl$ is the reduced Planck mass and we introduce the variable $x=\ms/T$ with the integration limits set by $T_R$ and $T_{\rm EWPT}$ for $x_{\rm min}$ and $x_{\rm max}$, respectively. The $K_i(x)$ functions are the modified Bessel functions of the $i^{\rm th}$ kind. We have also labelled this freeze-in yield with ${\rm UV}$ referring to the fact that the final flavon yield is driven by production close to the maximal temperature due to the non-renormalisable interactions. This is what is usually classed as ultraviolet freeze-in~\cite{Hall:2009bx}, and in our setup the maximal temperature is effectively identified with $T_R$. We also assume an instantaneous phase transition at $T_{\rm EWPT}=100~\gev$, which is sufficient due to the UV dependence of the yield. 

After the EWPT, one turns to the Lagrangian of \eqref{eq:Lafter}, where we see that the dimension 4, 3-point vertex will enable $2\rightarrow 1$ inverse decay from quarks into flavons. This is the dominant process as long as $\ms\geq (m_i + m_j)$, where $m_i$ are quark masses. When $\ms < (m_i + m_j)$, inverse decay is kinematically forbidden, then the $2\rightarrow 2$ scatterings such as $q\bar{q}\rightarrow \sigma g$ dominate. In this work, we will remain in the regime where $2\rightarrow 1$ production is always dominant. Since these processes are renormalisable, the majority of production happens at lower temperatures, i.e. in the infrared (IR) limit. The yield is given by 
\begin{align}
        Y_{ij}^{u/d,\,{\rm IR }}&\approx \frac{3 \left\vert \tilde{g}_{ij}^{u/d}\right\vert^2 A_{\star} \mpl}{16\pi^3 \ms}\left(1- 2\frac{m_i m_j}{\ms^2}\right) \nonumber\\
        & \times \sqrt{1-\frac{\left(m_i+m_j\right)^2}{m_{\sigma}^2}}\sqrt{1-\frac{\left(m_i-m_j\right)^2}{m_{\sigma}^2}}\int_{x_{\rm min}}^\infty {\rm d}x\, x^3 K_{1}(x), 
\label{eq:flavY_awept}
\end{align}
where $x_{\rm min}= \ms/T_{\rm EWPT}$ and one can safely take $x_{\rm max}\rightarrow \infty$ when keeping $\ms\gtrsim\mathcal{O}(\gev)$.\\

\begin{figure}
\includegraphics[width=0.48\textwidth]{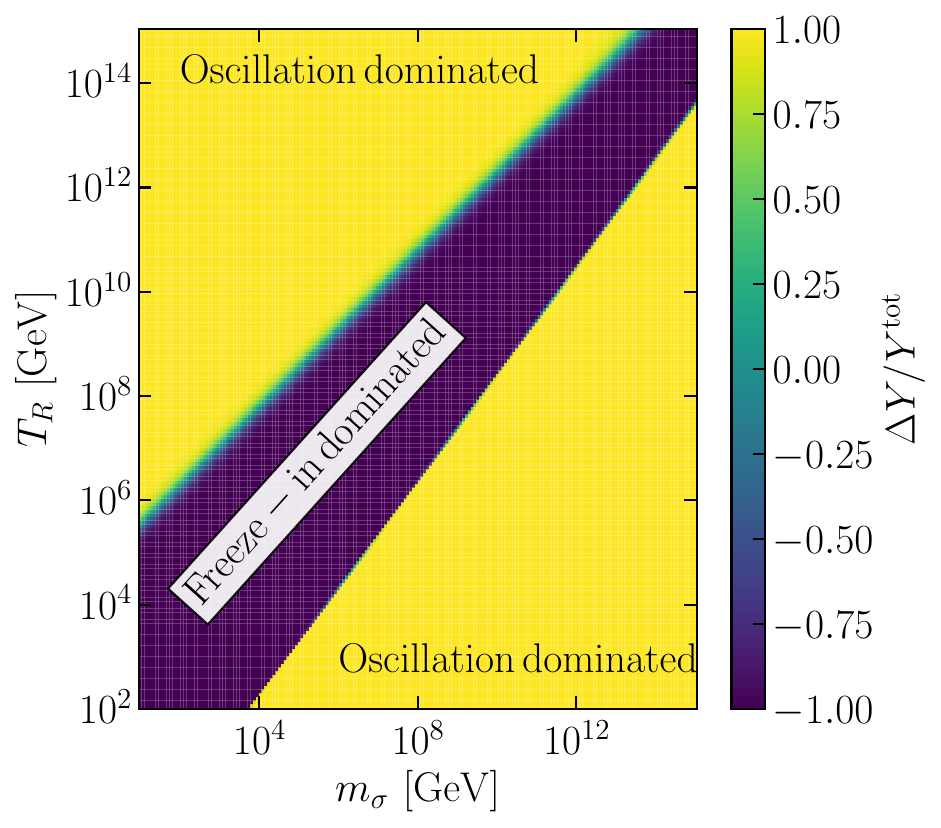}
\includegraphics[width=0.45\textwidth]{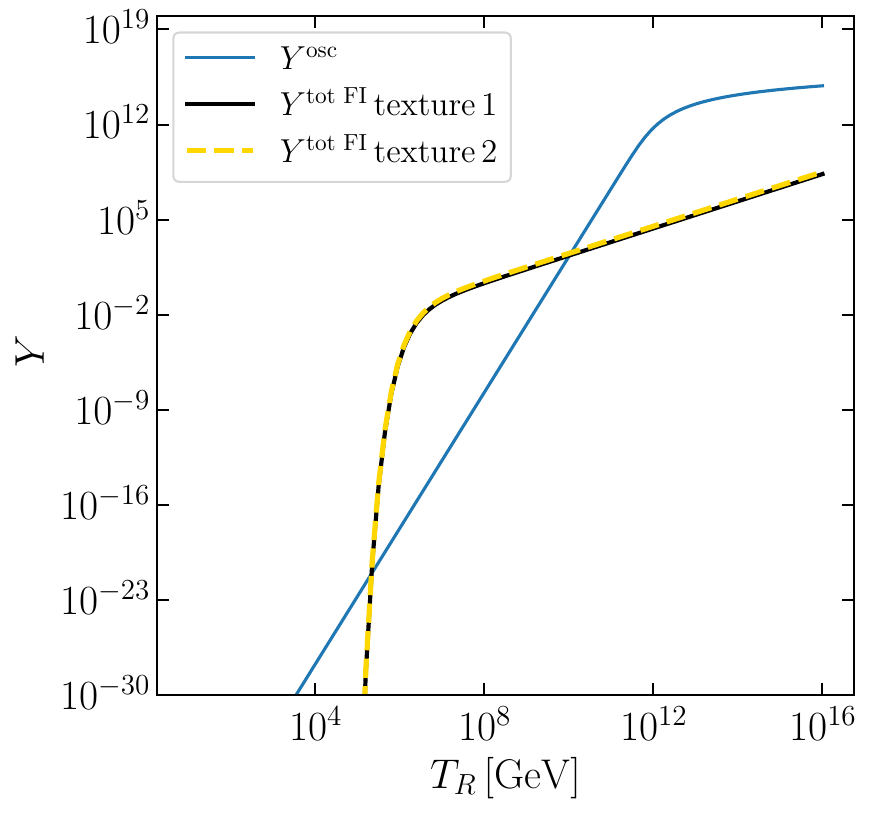}
\caption{Plot showing in which regions of parameter space the yield $Y$ is dominated by either freeze-in or oscillation production mechanism for the flavon. $\Delta Y = Y^{\rm osc.} - Y^{\rm FI.}$. \textbf{Left:} A heatmap where the yield is dominated by production from oscillations (yellow)  or freeze-in (dark blue). \textbf{Right:} The yield for oscillation production (blue) and freeze-in of texture 1 \Eqref{eq:charges1} (black solid) and texture 2 described in the paragraph below~\Eqref{eq:charges1} (yellow dashed). The flavon mass has been fixed to $m_\sigma=10^5~\gev$ and a range of $T_R$ values are taken. }
\label{fig:oscvsFI}
\end{figure}

For increasing $T_R$, the flavon production rate becomes increasingly dominated by a different mechanism based on flavon field oscillations. The flavon acquires a thermal potential due to thermal corrections to the free energy from the Yukawa couplings. This takes the form 
\begin{equation}
    V_{\rm eff} = \gamma T_{\rm Y}T^4 + \alpha T^4\frac{\sigma}{\Lambda} + \frac{m_{\sigma}^2}{2}\sigma^2 + \ldots
\label{eq:Veff}
\end{equation}
with \(\gamma = 5/96\), \(\alpha \approx 1/100\), and \(T_Y = {\rm tr}(Y_u^\dagger Y_u + Y_d^\dagger Y_d)\) \cite{Lillard:2018zts}. This shifts the minimum of the flavon potential and can result in flavon oscillation which behaves like matter. The relevant yield can then be obtained by solving its equation of motion. Assuming that the flavon initially occupies a minimum of a non-perturbed potential with $\sigma = 0$ and $\dot{\sigma}=0$, as well as working within the instantaneous-reheating approximation, one then obtains the following yield
\begin{equation}
Y^{\rm osc} \approx \alpha^2 \frac{A_\star \mpl \ms}{\Lambda^2} 
\left[\frac{M_H (T_R)}{\ms} \right]^{3/2} \times 
\begin{dcases}
\left( \frac{T_R}{T_*} \right)^{5} &~ \text{if}~T_R 
\lesssim T_*\;, \\ \Tstrut
2.1 \ln^2\left(\frac{T_R}{T_*} \right) &~\text{if}~ T_R \gg T_*
\; ,
\end{dcases}
\label{eq:Yosc}
\end{equation}
where $M_H\equiv \mpl/(0.33\sqrt{g_{\star}^{\rho}})$ and $T_\ast = \sqrt{m_\sigma M_{H}}$. As can be seen, when $T_R<T_\ast$, the flavon yield from the oscillation mechanism strongly grows with increasing $T_R$ and its energy-density can even dominate over the other components of the Universe. It is important to note that choosing different boundary conditions for the flavon field can substantially affect the final yield, especially for $T_R\gtrsim T_\ast$. However, as we will see below, in our case typically $T_R<T_\ast$ and the scaling of $Y^{\rm osc}$ with the reheating temperature is characterized by a behavior that is more universal.

In \cref{fig:oscvsFI} we see how the dominant production mechanism is determined by the parameters $T_R$ and $m_{\sigma}$. The colours of the heatmap depict $\Delta Y/Y^{\rm tot}$, where $\Delta Y=Y^{\rm osc}-Y^{\rm FI}$ and $Y^{\rm tot}= Y^{\rm osc}+Y^{\rm FI}$, such that the colormap saturates, i.e. is yellow (blue) for a value of $\Delta Y/Y^{\rm tot} = 1 \,\left(-1\right)$, when freeze-in production is 0\% (100\%) responsible for the flavon yield. Similarly, when  $\Delta Y/Y^{\rm tot} = 0$, both oscillation and freeze-in production contribute equally to the flavon abundance. There is no explicit reference to the value of $\Lambda$ chosen here, because all yields scale to the same power of $\Lambda$, therefore the relative efficiency of each process is independent of the cutoff. One can see a sharp contrast between regions of oscillation and freeze-in dominance. The oscillation mechanism generally dominates at high $T_R$ while for lower temperatures freeze-in becomes the most important production mode. Additionally, for $T_R\ll m_\sigma$, the freeze-in yield becomes highly suppressed again as heavy flavons can then only be produced in the high-energy tail of the momentum distribution of the SM species. Similar suppression is less pronounced for the oscillation mechanism which again becomes dominant. The right panel of \cref{fig:oscvsFI} shows, for a specific value of $m_\sigma$, how the behaviours of freeze-in and oscillation vary with $T_R$. We see that when $T_{R}\lesssim m_{\sigma}/10$, freeze-in production is highly suppressed. This is because the thermal plasma has insufficient energy to produce the flavon efficiently.

In this plot and the discussion below, we have focused on a specific choice of the FN charges (texture~1) \begin{equation}
\textrm{FN charges:\ \ }Q^{(1)}_{\rm FN}(\bar{Q}_{1,2,3}) = (3,2,0),\ \ Q^{(1)}_{\rm FN}(u_{1,2,3}) = (4,1,0),\ \ Q^{(1)}_{\rm FN}(d_{1,2,3}) = (3,2,2),
\label{eq:charges1}
\end{equation}
and follow the corresponding determination of the order-one coefficients $y_{ij}^{u/d}$ in \cref{eq:L} from Ref.~\cite{Lillard:2018zts}. Importantly, the freedom to choose the texture is limited in the SM sector by the requirement that it should replicate the observed mass hierarchy and the CKM matrix elements. We have also verified numerically that our results for the flavon yield from freeze-in do not change by more than a factor of a few for a different choice of a realistic texture following that study. We illustrate this in the right panel of \cref{fig:oscvsFI} where the yellow dashed line corresponds to the flavon freeze-in production for the following choice of the texture: $Q^{(2)}_{\rm FN}(\bar{Q}_{1,2,3}) = (3,2,0)$, $Q^{(2)}_{\rm FN}(u_{1,2,3}) = (5,2,0)$, $Q^{(2)}_{\rm FN}(d_{1,2,3}) = (4,3,3)$. We label this choice by texture 2 in the plot and compare it with the results obtained for the FN texture in \cref{eq:charges1} that is shown with the black solid line. As can be seen, varying the FN texture has a minor impact on the flavon yield and this would translate into only small changes in our results for the DM relic density that could, e.g., be absorbed in a slight change of $T_R$ as will be discussed below. In particular the difference in flavon yield from the two textures shown above only occurs for freeze-in production and is $Y^{{\rm FI}}\left({\rm texture \, 1}\right)\sim 0.7 \times Y^{{\rm FI}}\left({\rm texture \, 2}\right)$

Examining \cref{eq:flavY_bwept,eq:Yosc} one can appreciate why we observe such differences. Firstly, for oscillation production, \cref{eq:Yosc} shows no explicit dependence on FN charges of the SM species. The $\alpha$ parameter for example is determined by the potential and the quantity $T_{Y}$, cf. \cref{eq:Veff}. We take the latter to be fixed by the SM Yukawas~\cite{Lillard:2018zts} that should be reproduced by any given FN texture. Instead, for freeze-in, we note the dependence of the yield on the FN charges via $g_{ij}^{u/d}$ and $\tilde{g}_{ij}^{u/d}$ couplings in \cref{eq:flavY_bwept,eq:flavY_awept}. The difference of the corresponding flavon yields across textures is, however, expected to be small and not larger than an order of magnitude. We stress that flavon couplings to the SM species are dominant for the $2$nd and $3$rd generation quarks, especially for $t$, $b$, and $c$ quarks. For lighter quarks, they become suppressed by more powers of $\epsilon$. It is also customary to assign the FN charge $0$ to the top quark as its mass naturally lies close to the SM Higgs vev. The FN charges of both the bottom and charm quarks can then be assigned with only limited freedom, as they should reproduce the correct pattern of masses and CKM matrix elements relevant for these quarks. 

This translates into a limited hierarchy between the flavon couplings to these quarks. For UV freeze-in, \cref{eq:flavY_bwept}, differences between textures originate from the definition of $g_{ij}^{u/d}$ in \cref{eq:gbefore}. As a first-order check, we have taken the 15 FN charge configurations in table 2 of Ref~\cite{Fedele:2020fvh} and determined the $Y^{\rm UV}$ for each, assuming their $y_{ij}^{u/d}$ matrices are the same. We found that the range in yield was $\sim 6\%$ of the mean value only. Larger differences would arise after taking into account variations in $y_{ij}^{u/d}$ parameters for each of the textures. However, this should not impact the flavon yield by more than a factor of a few, as we have carefully verified for the two aforementioned textures. This is because the motivation behind the FN mechanism is to explain hierarchies while keeping fundamental parameters such as $y_{ij}^{u/d}$ of a similar order. Similarly, with IR freeze-in, texture differences are present only in the definition of $\tilde{g}_{ij}^{u/d}$ which are proportional to $n_{ij}^{u/d}$ and the relevant powers of $\epsilon$. While we leave a more comprehensive study of the impact of various FN textures on the flavon yield for the future, we argue this is not expected to significantly affect our conclusions.

Before we move on to the decay of the flavon into SM and dark sector particles alike, let us comment on flavon equilibration in the primordial plasma. In fact, we assume that the flavon is never in chemical equilibrium with the SM for high enough values of $\Lambda$ of our interest such that the aforementioned non-thermal production modes can dictate its final yield. To check whether this is the case, one determines whether or not the decoupling temperature $T_{\rm dec}$, defined by $H(T_{\rm dec})=n_X\langle\sigma v\rangle$, is above $T_R$. Approximately, 
\begin{equation}
    T_{\rm dec}\approx \frac{0.33\sqrt{g_{\star}^\rho}10^2\,\Lambda^2}{\mpl\sum_{ij}\left\vert g_{ij}^{u/d}\right\vert^2}\sim \Lambda\times\left(\frac{\Lambda}{10^{15}~\gev}\right),
    \label{eq:flavon_decoupled}
\end{equation}
and below we will verify if the condition $T_R<T_{\textrm{dec}}$ is satisfied. Notably, for $\Lambda<10^{15}~\gev$, this also implies that $T_R<\Lambda$ which allows us to rely on effective Lagrangians in \cref{eq:L,eq:Lchi1}. The requirement of being out of equilibrium is, however, more strict for lower values of $\Lambda$.

Let us also consider an extended period of reheating in which the inflaton dominates the energy density and decays to SM radiation at a constant rate. In such a more realistic, non-instantaneous reheating case, the maximum temperature of the Universe can grow above $T_R$ during the reheating period. The Hubble rate during such a period scales with temperature as \(H \propto T^4 / T_{\rm R}^2 M_{\rm P}\), while the freeze-in production rate of flavons scales as \(n_X\langle\sigma v\rangle \propto T^3 / \Lambda^2\). This implies that, if \(H > n_X\langle\sigma v\rangle\) at \(T = T_{\rm R}\), then freeze-in production is also inefficient throughout earlier reheating. The oscillation mechanism, however, is more involved and we leave the study of flavon production from oscillations during inflationary reheating for future work. In this work, we will assume that the flavon abundance does not reach equilibrium with the SM at earlier times and consider out-of-equilibrium production from freeze-in and oscillations during the radiation phase that follows reheating.

\subsection{Subsequent dark matter production}

Non-thermally produced flavons subsequently decay either back to the SM species or into lighter dark fermions $\chi_i$. The partial decay width of the flavon into two products, $f_i$ and $f_j$, is given by~\cite{Bauer:2016rxs},
\begin{align}
    \Gamma_{\sigma\rightarrow f_i f_j} &=\frac{\ms}{16\pi}\left[\frac{\left[\ms^2-(m_{f_i} + m_{f_j})^2\right]\left[\ms^2-(m_{f_i} - m_{f_j})^2\right]}{\ms^4}\right]\nonumber\\
    &\times\left\{\left(\left\vert   g_{ij}\right\vert^2+\left\vert g_{ji}\right\vert^2\right)\left(1-\frac{m_{f_i}^2+m_{f_j}^2}{\ms^2}\right)-4{\rm Re}(g_{ij}g_{ji})\frac{m_{f_i} m_{f_j}}{\ms^2}\right\}, 
    \label{eq:flavon_decay_general}
\end{align}
for $\ms>m_{f_i}+m_{f_j}$. The couplings $g_{ij}$ are generic couplings that can be replaced by the relevant coupling matrix, depending on what process one is interested in calculating. For example, the partial decay into dark sector particles, $\sigma\rightarrow \bar{\chi}_{i}\chi_j$, is given by 

\begin{equation}
    \Gamma_{\sigma\rightarrow \bar{\chi}_{i}\chi_j} = \Gamma_{\sigma\rightarrow f_i f_j}\left(g_{ij}\rightarrow g_{ij}^{\chi}\right),
\end{equation}
where we have substituted the coupling matrix defined in \Eqref{eq:flavon_chichi_int}. For flavon decays into SM particles, the decay processes depend on whether or not electroweak symmetry is broken. The two-body decay occurs if $\Gamma^{u/d,\,{\rm UV}}_{ij}< H(T_{\rm EWPT})$, and the partial width to up-type quarks is

\begin{equation}
    \Gamma^{u,\,{\rm IR}}_{ij}= N_c \Gamma_{\sigma\rightarrow f_i f_j}\left(g_{ij}\rightarrow \tilde{g}_{ij}^{u}\right)
\end{equation}
where $\tilde{g}_{ij}$ is defined in \eqref{eq:Lafter}. Similarly there is the partial width for down-type decay products. Alternatively, before EWPT, the 4-point vertex of \Eqref{eq:Lbefore} gives the 3-body decay, $\sigma\rightarrow \Phi\bar{Q}_i u_j$. The relevant decay width reads 
\begin{equation}
    \Gamma^{u/d,\,{\rm UV}}_{ij}=\frac{N_c}{3}\frac{\left\vert g_{ij}^{u/d}\right\vert^2}{64 \pi^3}\frac{\ms^3}{\Lambda^2},
    \label{eq:flavon_decay_SMUV}
\end{equation}
and it grows much more strongly with the flavon mass than both the visible IR and invisible decay widths. A useful temperature to define is the temperature of the SM plasma at time of flavon decay, $T_\sigma$, determined by $\Gamma_{\sigma}\simeq H(T_{\sigma})$. If $T_\sigma$ is above the EWPT, which we take to be $100~\gev$, the flavon decays visibly via \Eqref{eq:flavon_decay_SMUV}. In this regime, the dependence on $\ms$, means that for large flavon masses, its decay is dominantly into SM particles, making it harder to populate the Universe with dark fermions. Instead, if the flavon decay temperature $T_\sigma$ is lower than the one characteristic for the EWPT, DM production in late-time flavon decays can be effective. 

Once the flavon yield $Y_{\sigma}$ is produced as described above, the relic abundance of the $\chi_i$ species produced in flavon decays, $\Omega_{\chi_i} h^2$, is given by, 
\begin{equation}
    \Omega_{\chi_i} h^2 = \frac{m_{\chi_i}\,Y_{\chi_i}\,s_0h^2}{\rho_c}=m_{\chi_i}\,\gamma\,Y_{\sigma}\left(2\,{\rm Br}(\sigma\rightarrow \bar{\chi}_i\chi_i) + \sum_j{\rm Br}(\sigma\rightarrow \bar{\chi}_j\chi_i)\right)\frac{s_0 h^2}{\rho_c},
    \label{eq:relic}
\end{equation}
where $s_0$ and $\rho_c$ are the entropy density today and the critical energy density respectively. We include the parameter $\gamma$ as a dilution factor in the case of flavon domination, which is bounded by $\gamma\leq 1$. If flavons are so efficiently produced that they dominate over the other contributions to the energy density of the Universe, a short period of early matter domination ensues. Dilution occurs due to subsequent flavon decays that inject entropy into the SM plasma, temporarily altering its scaling due to cosmic expansion. The dilution factor is given by the ratio of the radiation entropy densities at $T_{\sigma}$ assuming no flavon, $s_{\rm before}$, and after flavon decays, $s_{\rm after}$, cf. e.g. Ref.~\cite{Ema:2018abj},
\begin{equation}
\gamma\equiv \frac{s_{\rm  before}}{s_{\rm after}} \simeq {\rm min}\left[\frac{3}{4}T_{\sigma}\frac{s^{\rm ini}}{\rho_\sigma^{\rm ini}}, 1\right]   ={\rm min}\left[\frac{3T_\sigma}{4\,\ms\,{\rm Br}(\sigma\rightarrow {\rm SM})\,Y_{\sigma}^{\rm ini}},1\right],    
\end{equation}
where $\rho_\sigma^{\rm ini}$, $s^{\rm ini}$ and $Y_\sigma^{\rm ini}$ are the flavon energy density, radiation entropy density and flavon yield once flavon production has ceased.

Similarly to visible flavon decays, in the dark sector the dominant decays are into the heaviest particles, i.e. those with the smallest FN charge. To illustrate the relative abundance of the lighter components in a multi-component dark sector, consider a two-component setup, $\chi_1$ and $\chi_2$ being the light and heavy respectively. As with the rest of this paper, we take FN charges of the dark sector particles to be left-right symmetric, $\mathcal{Q}_{\chi_{i,L}}=\mathcal{Q}_{\chi_{i,R}}$. Then $\Gamma_{\sigma\rightarrow\chi_1\chi_1}/\Gamma_{\sigma\rightarrow\chi_2\chi_2}\sim \epsilon^{4\Delta Q_{\chi}}$, where $\Delta Q_{\chi}=Q_{\chi_1}-Q_{\chi_2}$ and similarly $\Gamma_{\sigma\rightarrow\chi_1\chi_2}/\Gamma_{\sigma\rightarrow\chi_2\chi_2}\sim \epsilon^{2\Delta Q_{\chi}}$. In this case, the portion of the $\chi_1$ yield, $Y_{\chi_1}$ w.r.t. the combined dark sector yield $Y_{\rm DS}$ is
\begin{equation}
  \frac{Y_{\chi_1}}{Y_{\rm DS}}\approx\frac{2\epsilon^{4\Delta Q_{\chi}}+\epsilon^{2\Delta Q_{\chi}}}{2(\epsilon^{4\Delta Q_{\chi}}+\epsilon^{2\Delta Q_{\chi}}+1)},  
\end{equation}
and in the two-component scenario $Y_{\chi_2}=Y_{\rm DS}-Y_{\chi_1}$. The parameter of interest is the relic abundance $\Omega_{\rm DS}$ and the ratio of the two dark sector particles, $\Omega_{\chi_1}/\Omega_{\chi_2}$, which can be approximated as 
\begin{equation}
\frac{\Omega_{\chi_1}}{\Omega_{\chi_2}}\approx \epsilon^{2\Delta Q_\chi}\left(\frac{2\epsilon^{4\Delta Q_\chi}+\epsilon^{2\Delta Q_\chi}}{2-\epsilon^{2\Delta Q_\chi}}\right).
\label{eq:Omegaheuristic}
\end{equation}
Assuming only integer charges and taking $\Delta Q_{\chi}=1$, we observe that the subdominant $\chi_1$ contribution to the total DM relic abundance can achieve the largest portion of order $0.3\%$ of the $\chi_2$ relic density. This fraction is even smaller for increasing charge difference $\Delta Q_\chi$ and for lighter DM species in the multi-component scenario. For this reason, it is typically sufficient to focus on the heaviest dark sector fermion when calculating the DM relic abundance.

Importantly, the heavier dark sector particles in our setup can sequentially decay into the lighter ones semi-visibly, $\chi_{j}\to q\bar{q}\,\chi_i$, or even invisibly, $\chi_{i}\to \chi_j\,\chi_k\bar{\chi}_l$. In this case, if the lifetime of the heavier dark species is smaller than the age of the Universe, the final DM relic abundance today will correspond to the relic density of these lighter species. Again, in the simplest two-component DM scenario, the yield of $\chi_1$ will depend on the branching fractions into semi-visible and fully invisible final states 
\begin{equation}
Y_{\chi_1 \leftarrow \chi_2} = \left[\textrm{Br}(\chi_2\to q\bar{q}\,\chi_1) + 3\,\textrm{Br}(\chi_2\to 3\chi_1)\right]\times Y_{\chi_2}.
\end{equation}
This expression is valid in the non-relativistic regime of $\chi_1$. In such decays, however, lighter dark species are produced with relativistic boost factors given that $m_{\chi_2}\gg m_{\chi_1}$. For small redshifts at the time of $\chi_2$ decay, free-streaming constraints should then be taken into account in determining available regions in the parameter space of the model. This is discussed in section~\ref{sec:results} below. 

\section{Results and Dark matter phenomenology\label{sec:results}}

\subsection{One component dark sector\label{sec:onecomponentresults}}

In this section, we will discuss the general results of our calculations outlined in section \ref{sec:production}, namely the parameter values that conspire to give the correct relic abundance of dark matter for our model. As discussed above, we will focus on scenarios with high UV-completion scale, $\Lambda\gtrsim 10^8~\gev$, and on non-equilibrium flavon production relevant for $T_R\ll\Lambda$, as dictated by \cref{eq:flavon_decoupled}. We also allow the flavon mass to vary as an independent parameter with the lower bound $\ms\geq 10~\gev$. As we will show below, cosmological bounds on the model legitimize such a choice while it also allows us to focus on the quark sector of the SM and its couplings to the flavon. Furthermore, there is no a priori reason to limit the FN charge values of the dark matter, hence we explore the results for different values of $Q_{\chi}$.

As argued above, even in the case of a multi-component dark sector, flavon decays typically populate the Universe with dominantly one of the dark fermions $\chi_i$, specifically the heaviest one which remains kinematically accessible. In order to discuss the DM relic abundance from flavon decay it is then useful to first focus on a scenario with just a single DM component $\chi_1\equiv \chi$ which couples to the flavon according to \cref{eq:Lchi1} with the relevant dark FN charge $Q_\chi$. The latter determines the powers $n_\chi$, cf. \cref{eq:chargeschi}. In this case, the final relic abundance of $\chi$ is given by \cref{eq:relic} with only one $\chi$ particle, which depends on the flavon yield, $Y_\sigma$, and its invisible branching fraction, $\textrm{Br}(\sigma\to\bar{\chi}\chi)$.

\begin{figure}
\centering
\includegraphics[width=0.45\textwidth]{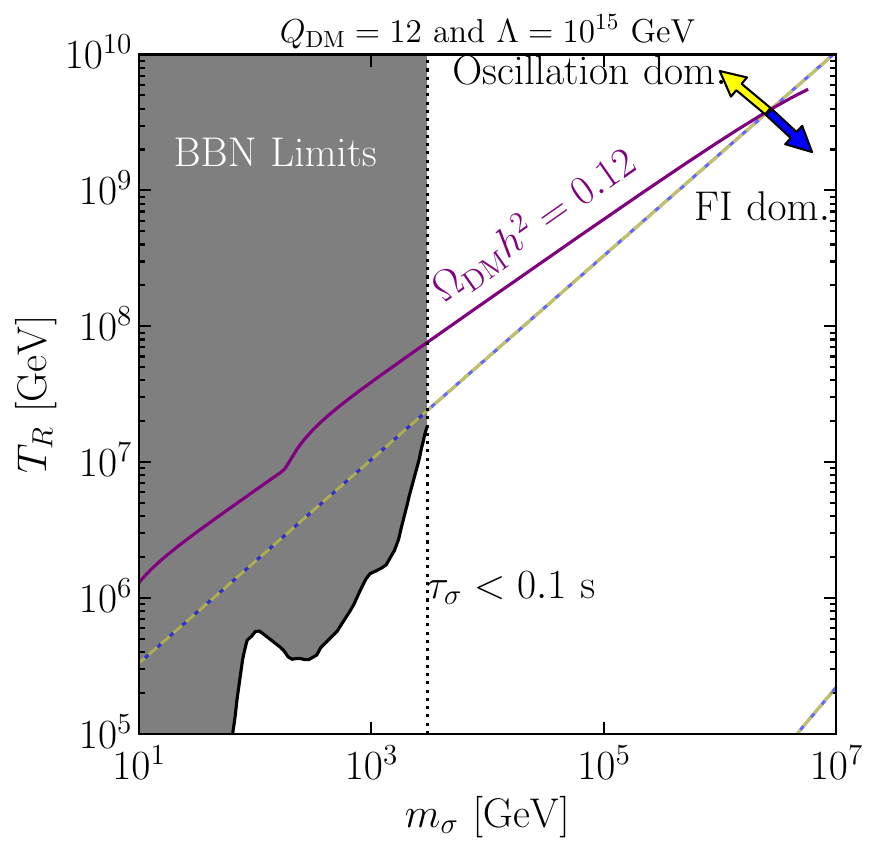}
\includegraphics[width=0.45\textwidth]{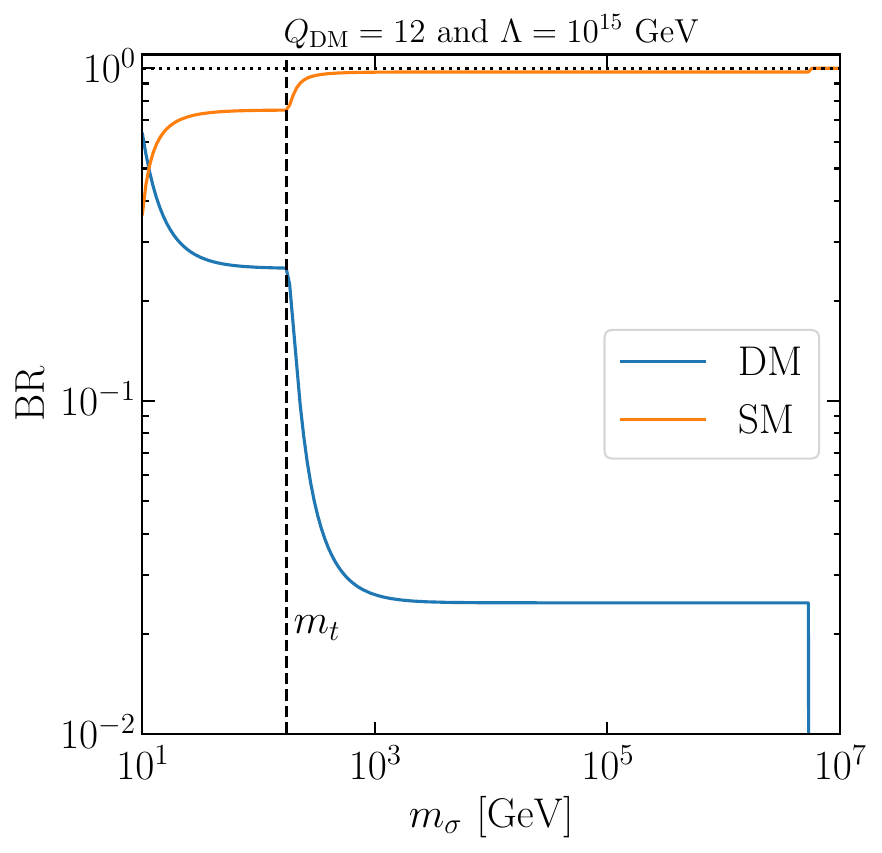}
\caption{\textbf{Left:} The BBN constraints (black shaded) on the $(m_\sigma, T_R)$ plane when there is also a dark Dirac fermion that couples to the flavon field. The purple line shows the points in parameter space that would produce the correct relic density for this dark matter particle. The vertical dotted line is the point where BBN limits are no longer relevant because the flavon decays before BBN begins. The blue and yellow stripped line indicates the transition from oscilation to freeze-in dominated production of flavons, as seen in Fig.~\ref{fig:oscvsFI}.  \textbf{Right:} Branching ratios for flavon decays into the dark and visible sectors in blue and orange, respectively. For both panels we show the example where $\Lambda=10^{15}~\gev$ and $Q_{\chi}=12$  that give $m_{\chi}=0.5~\gev$.}
\label{fig:BBN_lim_BRs}
\end{figure}

In the left panel of \cref{fig:BBN_lim_BRs}, we illustrate in the $(\ms,T_R)$ plane a DM relic density line along which $\Omega_\chi h^2\simeq 0.12$~\cite{Planck:2018vyg}. This has been obtained for $Q_{\textrm{DM}} = 12$ and for fixed value of $\Lambda = 10^{15}~\gev$, therefore the DM mass is also constant in the plot, $m_\chi = 0.48\,\gev $. For constant $\Lambda$ and a given value of $\ms$, the flavon yield is fully determined by $T_R$. Fixing these parameters and $Q_\chi$ also leads to a specific value of $\textrm{Br}(\sigma\to\bar{\chi}\chi)$. As a result, the relic target corresponds to a line in this plot. Below the line, the predicted DM relic density would be too low to explain the measured abundance while $\chi$ could be a subdominant DM component. Above the relic target line, $\Omega_\chi$ grows too large, and non-standard cosmological evolution would be required to avoid $\chi$ DM overproduction.

We indicate in the plot also, with a blue and yellow striped lines, the transition between the region in the parameter space for large $T_R$ where oscillation production of the flavon dominates and the region with freeze-in production, these transitions correspond to those shown in the left panel of Fig.~\ref{fig:oscvsFI}. One can see that the majority of the correct relic target line lies within the oscillation region. In this case, the decreasing value of the flavon mass enhances the flavon production rate, cf. \cref{eq:Yosc}, so that a lower reheating temperature is required to satisfy the DM relic density constraint. For $\ms\gtrsim 10^{6}~\gev$ we also see the region where the correct DM relic abundance is achieved with freeze-in flavon production. The flattening of the relic target line in this regime is due to a very mild dependence of $Y_\sigma$ on the flavon mass for the UV freeze-in production mechanism and for $\ms\ll T_R$. Simultaneously, we observe $\textrm{Br}(\sigma\to\chi\bar{\chi})\simeq 1$ in this case. Notably, for $\ms\gtrsim \textrm{a few}\times 10^6~\gev$, the transition to very short-lived flavons occurs. These decay before EWPT via three-body decays into the SM species, cf. \cref{eq:flavon_decay_SMUV}, while the invisible branching fraction becomes completely negligible. As a result, the relic target line suddenly cuts there and we find no points with $\Omega_\chi h^2\simeq 0.12$ for such heavy flavons.

For decreasing $\ms\lesssim 10~\tev$, the flavon lifetime and its visible branching fraction grow. This is, eventually, constrained by cosmological bounds from Big Bang Nucleosynthesis (BBN). In order to determine whether a flavon-DM configuration is constrained this way, we consult the results of Ref.~\cite{Kawasaki:2017bqm}, where the energy injection from late time decays of heavy particles into SM species can subtly effect the properties of the thermal bath and disrupt the delicate balance that occurs at BBN. Ref.~\cite{Kawasaki:2017bqm} presents their findings as constraints on specific decay channels. The most relevant for our purposes are the $\bar{t}t$, $\bar{b}b$ and $\bar{u}u$ channels. We find that the BBN constraints can be avoided in our model if
\begin{equation}
m_{\sigma}Y_{\sigma}\left(T_R, m_\sigma, \Lambda\right) < \sum_{\bar{q}q} \frac{\Gamma_{\bar{q}q}\left(Q_{\rm DM}, \Lambda, 	m_\sigma \right)}{{\rm UL}_{\bar{q}q}\left(m_{\sigma}, \tau_\sigma\right)},
\end{equation}
where ${\rm UL}_{\bar{q}q}$ is the limit taken from Ref.~\cite{Kawasaki:2017bqm}. We have kept the parameter dependencies of each component for clarity and assume that all light ($m_q\leq m_{s}$), medium ($m_s<m_q\leq m_{b}$) and heavy ($m_q=m_{t}$) quark decay channels have similar limits corresponding to $\bar{u}u$, $\bar{b}b$, or $\bar{t}t$ channel, respectively. The resulting bounds are shown in \cref{fig:BBN_lim_BRs} as a grey-shaded region. As can be seen, for the specific benchmark values of the model parameters discussed above, the BBN constraints tend to exclude regions in the parameter space with $\ms \lesssim 2~\tev$. Heavier flavons in the plot have lifetimes $\tau_\sigma\gtrsim 0.1~\textrm{s}$ and are, therefore, not affected by the cosmological bounds.

The right panel of \cref{fig:BBN_lim_BRs} illustrates the dependence of the visible and invisible flavon branching fractions on $\ms$ for the same benchmark scenario. As can be seen, for very large flavon masses, $\ms\gtrsim \textrm{a few}\times 10^7~\gev$, flavon decays are fully dominated by the SM final states. In this regime, flavon decays before the EWPT with a large value of the SM decay width. For lower $\ms$, the invisible branching fraction grows up to $\sim 2~\%$. In this case, since $m_\chi< m_t$ and the flavon couples more strongly to the SM species, the SM decays with a top quark in the final state still dominate until the kinematic threshold for the $\sigma\to t\bar{t}$ and $t\bar{c}$ processes. For lower $\ms$, we find larger values of $\textrm{Br}(\sigma\to\chi\bar{\chi})$ which further grows close to the threshold for the bottom quark in the final state. 

\begin{figure}
\centering
\includegraphics[width=0.45\textwidth]{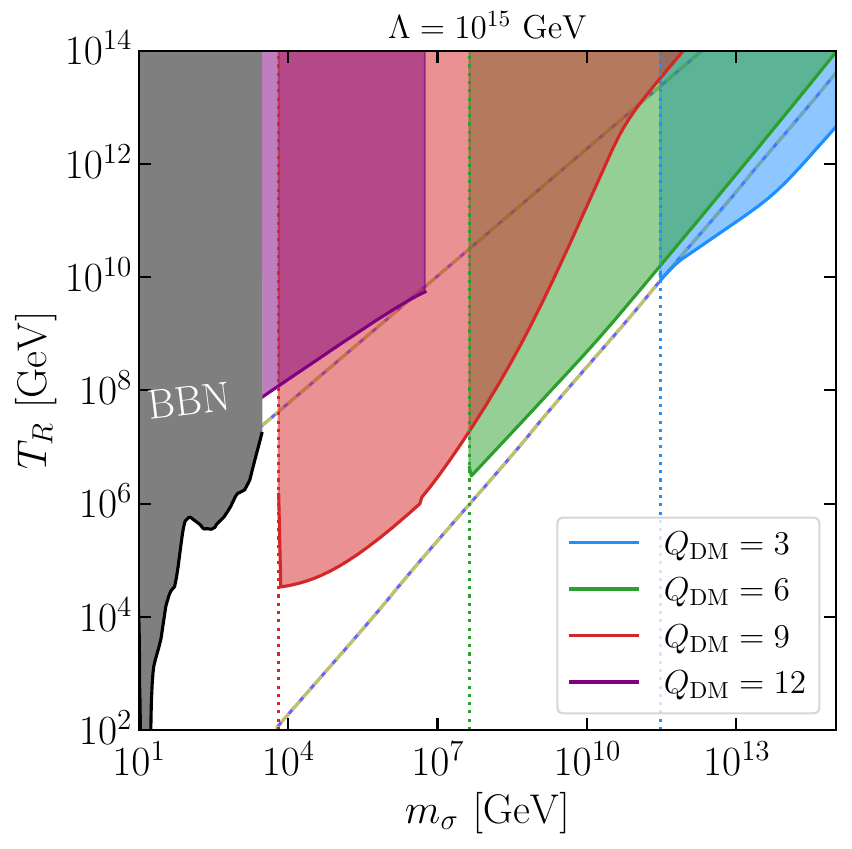}
\includegraphics[width=0.45\textwidth]{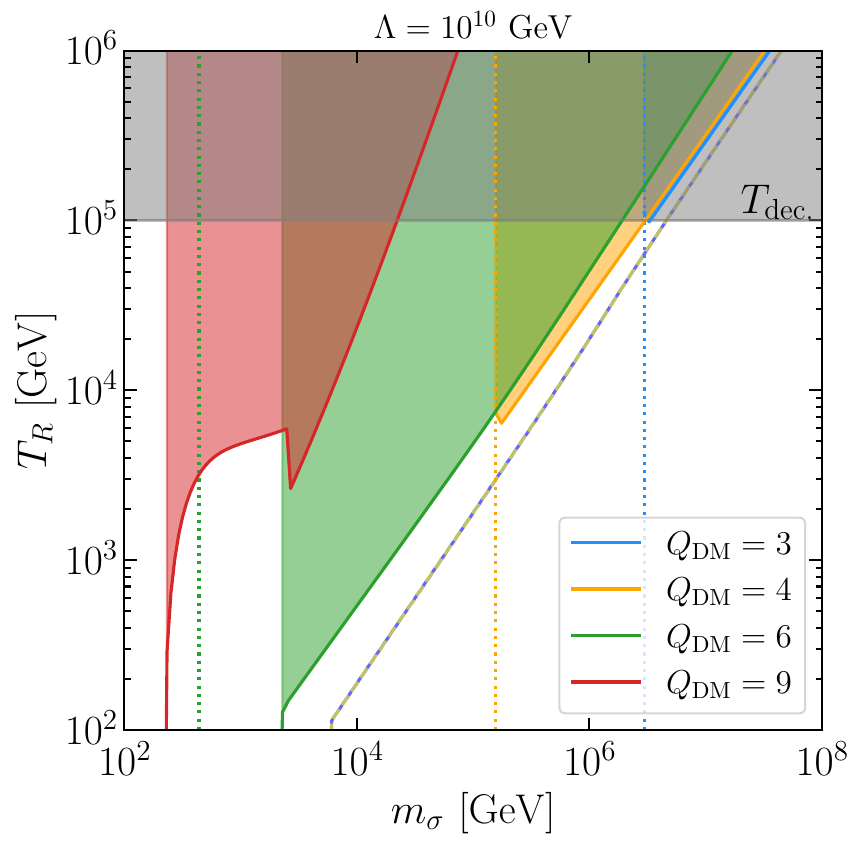}
\caption{Lines showing the values of $T_R$ and $m_{\sigma}$ needed to produce the correct DM relic density for charges 3 (blue), 4 (orange), 6 (green), 9 (red), 12 (purple). The lines are taken after BBN constraints have been taken into account. \textbf{Left:} $\Lambda=10^{15}~\gev$ and \textbf{right:} the same but for $\Lambda=10^{10}~\gev$. We have now shaded the regions where $\Omega_{\rm DM}h^2 \geq 0.12$.}
\label{fig:relic_lines}
\end{figure}

Figure~\ref{fig:relic_lines} shows the correct relic lines for a range of charge choices. We have taken two representative values for $\Lambda$ here, $10^{15}~\gev$ and $10^{10}~\gev$, as well as several choices of the dark FN charges, $Q_\chi = 3, 6, 9,$ and $12$. The larger the charge, the lower the DM mass which is equal to $m_\chi = 1.5\times 10^{11\,(6)}~\gev,\,2.2\times 10^{7\,(2)}~\gev,\, 3.2\times 10^{3\,(-2)}~\gev,\, 4.8\times 10^{-1\,(-6)}~\gev$ in the plot for $\Lambda = 10^{15}~\gev$ ($10^{10}~\gev$), for each of these benchmark $Q_\chi$ values, respectively. We also show, as before, the lines that separate regions in the parameter space of the model corresponding to dominant flavon production due to freeze-in and oscillation, cf. \cref{fig:oscvsFI}. In the right panel, the grey-shaded region on top indicates the regime of high $T_R$ in which the flavon is expected to equilibrate in the early Universe. 
The dotted vertical lines, the kinematic limits for $\ms$ below which flavons can no longer decay into $\chi\bar{\chi}$ pairs. These mark the low-mass ends of the relevant relic target lines in the left panel obtained for $\Lambda = 10^{15}~\gev$. The only exception is the $Q_\chi = 12$ case discussed above, which is constrained by the BBN bounds for lighter flavons. We stress that these bounds would be different for other values of $Q_\chi$. These other BBN bounds, however, are not shown in the plot as it is not relevant for the corresponding relic target lines. Instead, in the right panel where $\Lambda = 10^{10}~\gev$, the DM relic density for charges $Q_{\textrm{DM}} = 6$ and $9$ (red and green shaded regions) have their lower value of a valid $\ms$ determined by a different effect. One can see that the relic line extends almost vertically downward for at such a minimum $\ms$, but this value is larger than the kinematical threshold. This is driven by the transition from UV to IR freeze-in production at low $T_R$ since the latter flavon yield is largely independent of the precise value of the reheating temperature. In the $Q_{\textrm{DM}} = 9$ case, an additional drop of the relic target line for $\ms\sim \tev$ is due to the increase of the invisible decay branching fraction of the flavon when its mass approaches the threshold for decays into the top quarks.

In the limit of large flavon mass, the specific behavior of each relic line is determined by a number of factors. On the left panel, for $Q_{\chi}=12$ and $Q_{\textrm{DM}}=3$ we see the relic lines vary as they move from regions where flavons are dominantly produced by oscillations or freeze-in. We see that in the $Q_\chi=6$ case, a fairly straightforward freeze-in production scenario is exhibited, where, as $m_\sigma$ increases, a greater $T_R$ is required to produce the correct relic abundance with $\Omega_\chi \propto Y_\sigma^{\textrm{UV}}\propto T_R$ for $T_{\textrm{EWPT}}\ll \ms\ll T_R$~\cite{Lillard:2018zts}. This behavior is modified for $Q_{\textrm{DM}} = 9$ in the left panel. Here, for growing $\ms$ we enter the regime in which flavon decays happen before the EWPT. The increase of the flavon decay width to the SM species, in this case, results in a relative decrease of the invisible branching fraction. Therefore, a larger $T_R$ is needed to obtain the correct DM relic density. The shape of the relic target line changes again close to the region where the dominant flavon production is due to oscillations. A strong dependence of $Y_\sigma^{\textrm{osc}}$ on $T_R$ leads in this case to a less steep line in the $(\ms,T_R)$ plane, i.e., here even small changes of $T_R$ significantly affect $\Omega_\chi$ for increasing $\ms$. The same behavior can be observed for $Q_{\textrm{DM}}=3$ for $\Lambda = 10^{15}~\gev$ (blue-shaded region). Interestingly, in this scenario, the correct DM relic abundance can be obtained for $\ms\sim v_s$ without the need of a large fine-tuning of the parameters in the flavon potential. Similar can be observed for even larger dark FN charge $Q_{\textrm{DM}} = 6$.

As stressed above and shown in the left panel on \cref{fig:relic_lines}, for increasing dark FN charge the relic target lines become constrained by the BBN bounds. This effectively introduces an upper limit on the dark FN charge up to which one can obtain the correct DM relic density from flavon decays. In particular, for $\Lambda = 10^{15}~\gev$ we found that for $Q_{\textrm{DM}} = 13$ the entire relic target line is excluded this way for $\ms\gtrsim 10~\gev$. While these bounds could potentially be avoided for much lighter flavons, $\ms\ll\gev$, this would correspond to the regime of extreme fine-tuning with the flavon mass being many orders of magnitude suppressed with respect to its vev. A similar effect is true for lower $\Lambda$, as shown in the right panel. While we found that for the $Q_{\textrm{DM}} = 10$ scenario lies fully within the flavon equilibration regime, for even larger dark FN charges, e.g. $Q_{\textrm{DM}} = 12$, we also found strong BBN bounds affecting this scenario. It is interesting to comment on the scenario with the UV-completion scale set at the Planck scale, $\Lambda = M_P$. In this case, we found strong BBN bounds excluding models with $Q_{\textrm{DM}}\gtrsim 13$.

\subsection{Multi-component dark sectors and indirect detection\label{sec:multicomponentresults}}

While, as argued above, the DM relic abundance from flavon decays can effectively be discussed by focusing on just a single dark fermion $\chi_i$, further interesting phenomenological implications arise when multi-component dark sectors are considered. Here we discuss such implications and we reiterate that we have chosen the left-right symmetric charge configuration. We can therefore, use the relic-abundance results from \cref{sec:onecomponentresults}. 

Once the heavier dark fermion has been produced from flavon decays, it can itself decay into the lighter dark-sector particles via $\chi_i\rightarrow \chi_j \bar{q}q$ or $\chi_i\rightarrow \chi_j \bar{\chi}_k\chi_l$ where $i>j,k,l$ in our notation. Depending on the lifetime of the dark fermions, the DM content of the Universe will either be dominated by the heaviest fermion or by one of the lighter dark species produced in decays of $\chi_i$. The heaviest dark fermion is the dominating $\OmegaDM$ contribution when $\tau_{\chi_i}$ exceeds the age of the Universe, while for the lifetimes much lower than the recombination time, $\tau_{\chi_i} \ll \tau_{\textrm{recomb.}}$, the DM particle will be some lighter member of the dark sector. Scenarios with intermediate lifetimes are strongly bounded by cosmological constraints. 

By inspecting the interaction terms of \eqref{eq:flavon_chichi_int} and neglecting additional small phase-space corrections in the regime of $\ms\gg m_{\chi_i}\gg m_{\chi_{j,k,l}}$, we can approximate the dark decay width as 
\begin{equation}
    \Gamma_{\chi_i\rightarrow\chi_j\bar{\chi}_k\chi_l}\approx\frac{m_{\chi_i}^5}{512\pi^3m_\sigma^4} \epsilon^{2n^{\chi}_{ij}}\left(n^\chi_{ij}+1\right)^2\epsilon^{2n^{\chi}_{kl}}\left(n^\chi_{kl}+1\right)^2.
    \label{eq:dark_3body}
\end{equation}
As can be seen, decay widths into lighter dark fermions that are characterized by larger values of $n^\chi_{ij}$ and $n^\chi_{kl}$ are suppressed by additional powers of $\epsilon$. As a result, the dominant invisible decay mode of $\chi_i$ is into the next heaviest dark fermions, $\chi_i\to \chi_{i-1}\,\bar{\chi}_{i-1}\,\chi_{i-1}$, which is always kinematically possible if they differ by at least one unit of the dark charge. This decay mode competes with the semi-visible decay channel into quarks in the final state
\begin{equation}
    \Gamma_{\chi_i\rightarrow\chi_j\bar{q}_k q_l}\approx\frac{m_{\chi_i}^5}{512\pi^3m_\sigma^4} \epsilon^{2n^{\chi}_{ij}}\left(n^\chi_{ij}+1\right)^2\left(\tilde{g}_{kl}^{u/d}\right)^2.
    \label{eq:semivisible}
\end{equation}
Since our focus below will primarily be on late-time decays of $\chi_i$, we employ the expression above which is valid after the EWPT and employs the relevant couplings $\tilde{g}_{kl}^{u/d}$ of quarks to the flavon which scale as $v_{\rm EW}/2 \Lambda$. This decay mode is also dominantly into the next heaviest dark fermion, $\chi_i\to \chi_{i-1}\,\bar{q}_k\,q_l$. For the SM texture that we use in this study \cref{eq:charges1}, the largest value of the quark-flavon coupling comes from $\tilde{g}^{d}_{33}\sim 0.1 \,v_{\rm EW}/ \Lambda$. With the two expressions above and this value for $\tilde{g}^{d}_{33}$, we see that the semi-visible decays dominate over dark decays when
\begin{equation}
    \Lambda \lesssim \frac{v_{\rm EW}}{\epsilon^{n^{\chi}_{kl}}\left(n_{kl}^{\chi}+1\right)},
    \label{eq:DSSM_flip}
\end{equation}
where the most important is the case of $k = l = i-1$. In particular, for the growing dark charge $Q_{\chi_{i-1}}$ semi-visible modes can dominate for increasingly large values of $\Lambda$. However, there is always this turnover for where the dark decays dominate. The heavier dark fermions then preferentially decay invisibly into the lighter ones in a sequential way, schematically $\chi_i\to \chi_{i-1}\to\chi_{i-2}\to\ldots$, while even small semi-visible branching fractions can be bounded by astrophysics and cosmology, as we discuss below.

As can be seen from \eqref{eq:dark_3body}, the ratio of the dark particle lifetimes can be approximated by
\begin{equation}
    \frac{\tau_{\chi_i}}{\tau_{\chi_{i-1}}}\approx\left(\frac{m_{\chi_{i-1}}}{m_{\chi_{i}}}\right)^5 \epsilon^{2\left(n_{i-1,i-2}^{\chi}+n_{i-2,i-2}^{\chi}-n_{i,i-1}^{\chi}-n_{i-1,i-1}^{\chi}\right)}=\epsilon^{6\left(Q_{\chi_{i-2}} + Q_{\chi_{i-1}} - 2\,Q_{\chi_{i}}\right)}\,.
    \label{eq:tau_ratios}
\end{equation}
Remembering that $Q_{\chi_{i-2}}\geq Q_{\chi_{i-1}} \geq Q_{\chi_{i}}$, the ratio above is always less than 1, and typically $\tau_{\chi_i}/\tau_{\chi_{i-1}}\ll 1$.\footnote{This is the case unless all three charges are identical but dark decays will then typically be kinematically forbidden.} For example, with the three-component configuration $(Q_{\chi_1},\,Q_{\chi_2},\,Q_{\chi_3})=(3,2,1)$, one obtains $\tau_{\chi_i}/\tau_{\chi_{i-1}}\sim\epsilon^{18}\lesssim 10^{-11}$. The dark sector dynamics will usually be dominated at any given time by the decay of one dark species into the next lightest one and to quarks. For this reason, the essential aspects of the phenomenology of these models can be captured by focusing on a two-component scenario while a discussion of the full multi-component model can be largely reproduced as a sequence of such two-component cases. For the rest of this section we will focus on the two $\chi_i$ scenario. The mass hierarchy between them, $m_{\chi_2}>m_{\chi_1}$, is dictated by the respective dark FN charges, which we will vary below. 

In particular, the charge of the heavier dark fermion should be chosen such that it allows for obtaining correct DM relic density from flavon decays, as discussed in \cref{sec:onecomponentresults}. This is crucial for large $\chi_2$ lifetimes exceeding the age of the Universe. Instead, for small $\chi_2$ lifetimes, one should additionally take into account its subsequent decays producing $\chi_1$ species prior to recombination such that the observed $\chi_1$ DM relic abundance is bound to match the Planck observations~\cite{Planck:2018vyg}. The DM relic abundance in this case is then given by
\begin{equation}
\Omega_{\rm DM}\,h^2 =
    \begin{cases}
          2\,m_{\chi_2}\gamma\,Y_{\sigma}\,{\rm Br}(\sigma\rightarrow \bar{\chi}_2\chi_2)\frac{s_0 h^2}{\rho_c} &\text{if } \tau_{\chi_2} \gg \tau_{\rm rec.}\\
          2\,m_{\chi_1}\gamma\,Y_{\sigma}\,{\rm Br}(\sigma\rightarrow \bar{\chi}_2\chi_2)\times\left[{\rm Br}(\chi_2\rightarrow \chi_1\bar{q}q)+3{\rm Br}(\chi_2\rightarrow \chi_1\bar{\chi}_1\chi_1)\right]\frac{s_0 h^2}{\rho_c}&\text{if } \tau_{\chi_2} \ll \tau_{\rm rec.}
    \end{cases}
    \label{eq:relic_cases}
\end{equation}
Late-time decays of $\chi_2$ also lead to further constraints on this model. When semi-visible decay branching fraction is significant, the $\chi_2$ decay may be detectable or already constrained by indirect probes for decaying DM particles and cosmology. In \cref{sec:onecomponentresults}, we have already seen the implications flavon decay can have on BBN, the same is true for $\chi_2$ decay with quarks in the final state, and we apply the constraints from Ref.~\cite{Kawasaki:2017bqm} in a similar way to what was described in the previous section. Where we approximate that the $\bar{q}q$ products in the 3-body decay take $50\%$ of the centre of mass energy. We do this because the results in Ref.~\cite{Kawasaki:2017bqm} are specifically for 2-body decays into SM particles, and performing a full analysis which takes into account the full distribution of SM particles and their effects on BBN is beyond the scope of this paper and unlikely to affect our results in an appreciable way.

In addition to BBN constraints, there are now indirect signals coming from continua of $\gamma$-rays and neutrinos that would be observable by current and future telescopes~\cite{Cohen:2016uyg, Chianese:2021jke, Anchordoqui:2021crl, Hutten:2022hud, Bartlett:2022ztj}. The studies that provide the most stringent constraints are that of Fermi-LAT and IceCube as reported in Ref. \cite{Cohen:2016uyg} and the ultra high energy cosmic ray (UHECR) bounds from~\cite{Chianese:2021jke}. Additionally, energy injections into the thermal SM plasma of the early Universe can affect the spectral distortions and anisotropies of the Cosmic Microwave Background (CMB), limits on decaying dark matter have been comprehensively studied in Ref.~\cite{Lucca:2019rxf}  based on the data collected by the Planck~\cite{Planck:2018vyg} and COBE/FIRAS~\cite{Fixsen:1996nj} collaborations. In order to take the limits from Ref~\cite{Lucca:2019rxf} (Figure 8), we interpret the dark matter fraction $f_{\rm DM}$ as $\sum_{kl}{\rm Br}(\chi_2\rightarrow\chi_1\bar{q}_k q_l)$, with an additional scaling of 5 which we include to account for the fact that our hadronic final states do not affect the CMB with 100\% efficiency \cite{Slatyer:2015jla}.

For the case where $\chi_1$ is our dark matter particle, its free-streaming length will determine the minimum scale of structure in the matter power spectrum. This is due to the mass hierarchy between $\chi_2$ and $\chi_1$, such that the decay of $\chi_2$  produces boosted $\chi_1$ particles. The smallest scales in the matter power spectrum that can be experimentally probed are currently done so by studying the Lyman-$\alpha$ forest~\cite{Palanque-Delabrouille:2013gaa,2016A&A...594A..91L,Viel:2013fqw}. In the paradigm of cold dark matter, the free-streaming length is essentially zero. The boosted $\chi_1$ can have a free-streaming length sufficient to wash out small-scale structure. To determine whether certain parameter configurations are incompatible with results from Lyman-$\alpha$, we follow the methodology of Ref.~\cite{Garny:2018ali} (section D). 

\begin{figure}
    \centering
    \includegraphics[width=0.45\textwidth]{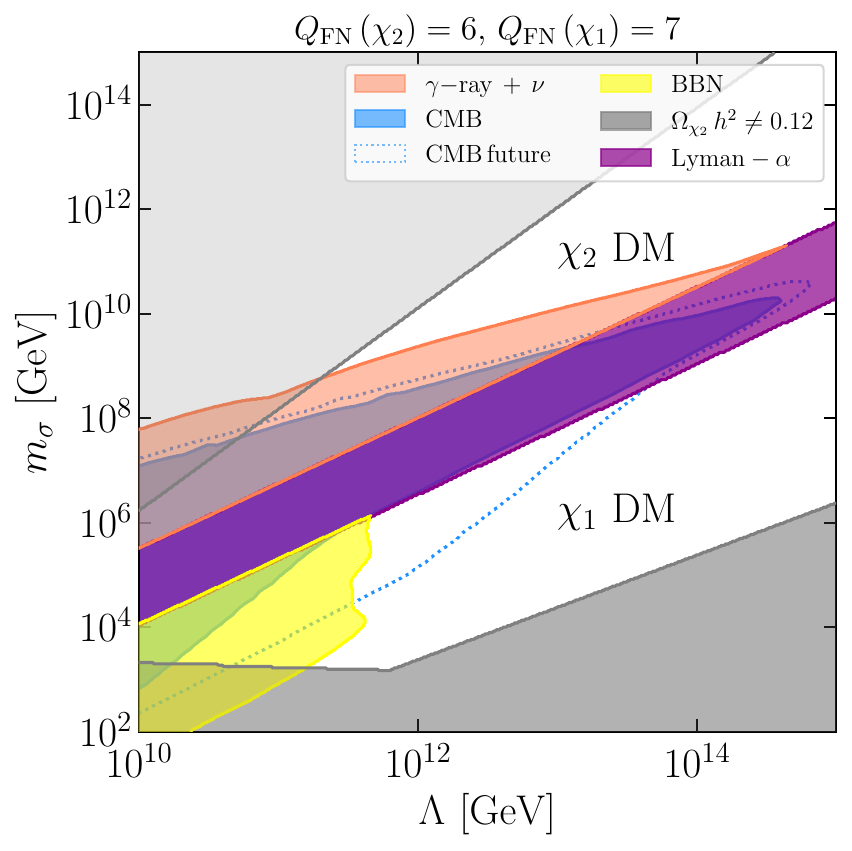}
    \includegraphics[width=0.45\textwidth]{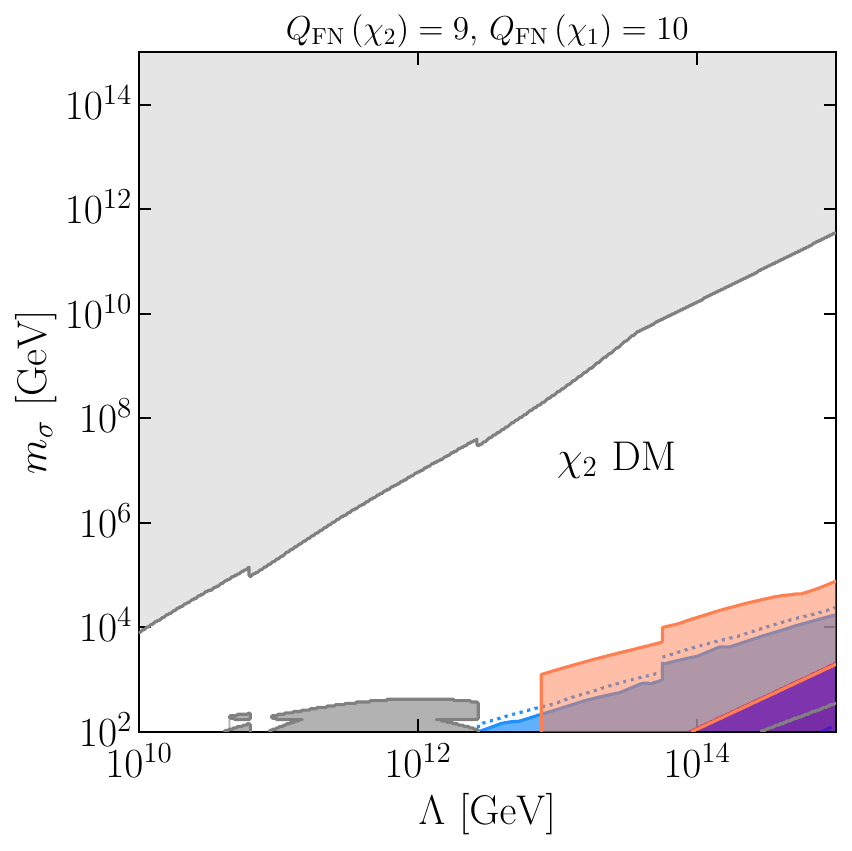}\\
    \includegraphics[width=0.45\textwidth]{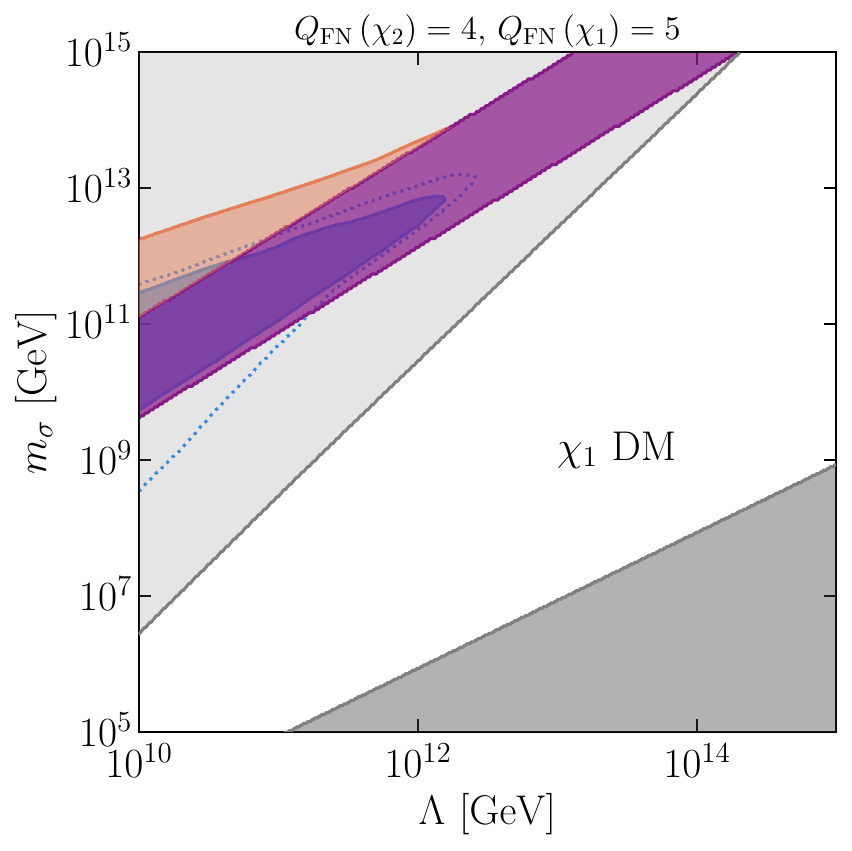}
    \includegraphics[width=0.45\textwidth]{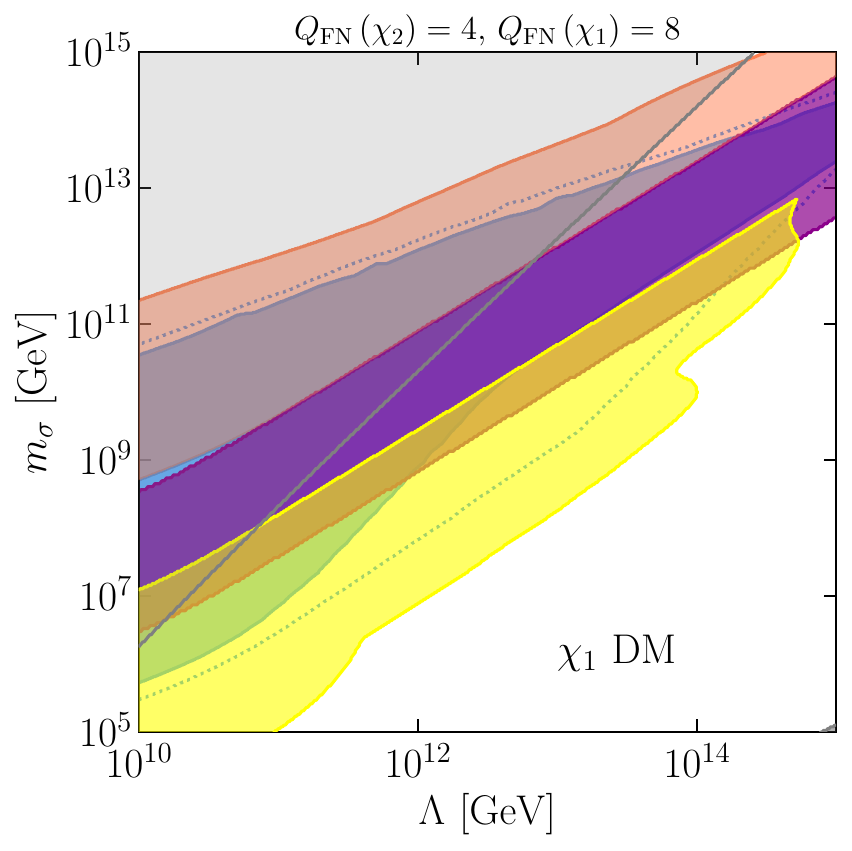}

    \caption{Parameter space for two-component flavon dark matter for a number of charge configurations. Shaded regions show constrained or not viable regions. The grey regions outline where the correct relic abundance is not achievable for any $T_R$ between $10^2~\gev$ and $T_{\rm dec}$. Constraints for decaying dark matter into visible signals are marked by coral, yellow and blue, for $\gamma$-ray and neutrino searches, BBN and measurements of the CMB respectively. Warm DM constraints from Lyman-$\alpha$ is shown in purple. Future probes of the CMB are projected with a blue dotted line. The white region is viable for a consistent description of dark matter. We have denoted on each panel which fermion, $\chi_1$ or $\chi_2$ is the dominant dark matter particle. }
    \label{fig:multi_ID}
\end{figure}

Figure~\ref{fig:multi_ID} shows how these indirect constraints manifest for particular charge choices of the two-particle dark sector. We have chosen to visualise this in the $(\Lambda, m_{\sigma})$ plane, where the charges of both $\chi_1$ and $\chi_2$ dictate their mass for a given $\Lambda$, \eqref{eq:mdm_eq}. Additionally, for these plots, we have not fixed a value for $T_{R}$, instead we have performed a numerical check to ensure that the $T_R$ required to produce the correct relic abundance lies between $10^2\,{\rm GeV}$ and $T_{\rm dec}$, where $T_{\rm dec}$ is the $\Lambda$-dependent temperature that determines whether the flavon has remained outside equilibrium throughout, see \eqref{eq:flavon_decoupled}. We show regions of parameter space that are unable to satisfy this condition in grey. As exemplified in the previous section, being unable to produce the correct relic abundance for any valid $T_R$ could be for a number of reasons. The upper bound could come from the ${\rm BR}\left(\sigma\rightarrow \bar{\chi}\chi\right)\rightarrow 0$ as with $Q_{\rm DM}=12$ (see Fig.~\ref{fig:BBN_lim_BRs}) or simply that the required $T_R$ moves above the decoupling temperature $T_{\rm dec}$ (see Fig.~\ref{fig:relic_lines} right panel). In the latter case, the correct DM relic abundance could still be reproduced for certain values of $\ms$ and $\Lambda$ in the thermal-production regime which is, however, beyond our focus in this study. We, therefore, highlight these regions in the parameter space with a light gray-shaded color. All upper bounds in Fig.~\ref{fig:multi_ID}, the behaviour is determined by $T_{\rm dec}$. Lower bounds can be set by the kinematic limit, where the flavon is unable to decay into dark matter, or by the interplay of UV and IR freeze-in as discussed for $Q_{\rm DM}=6$ and $9$ (see Fig.~\ref{fig:relic_lines}). The kinematic limit dictates the lower boundary when the limit follows a $\ms\propto \Lambda$, as can be seen for the whole bottom left panel. Note that this limit now corresponds to the $\sigma\rightarrow \bar{\chi}_1\chi$ limit. The flatter and more complicated lower bounds seen in the top row of Fig.~\ref{fig:multi_ID} are due to the IR and UV freeze in effect.

The dark charge configurations we have chosen to show are representative of a few different effects, all of which center around exhibiting the impact indirect detection and cosmological bounds have on these models. We show the existing indirect constraints from $\gamma$-rays and cosmic neutrinos taken from Refs.~\cite{Cohen:2016uyg, Chianese:2021jke} in shaded coral. The existing and future CMB sensitivities are displayed via the shaded blue region and the dotted blue lines respectively. Portions of parameter space where Lyman-$\alpha$ measurements are prohibitive are shown via the shaded purple regions. The yellow shaded region signifies the areas where $\chi_2\rightarrow \chi_1 \bar{q} q$ disrupts BBN.

The top left panel showing the charge assignment of $Q_{\rm FN}(\chi_{2})=6$ and $Q_{\rm FN}(\chi_1)=7$, exhibits an example where there are substantial values of $m_{\sigma}$ and $\Lambda$ that are already restricted by existing constraints. All four probes in this case give some degree of complementarity, where only the current CMB limits are completely covered by other searches. Note, however, that the future projections for CMB probes will venture into uncharted areas of parameter space. One can infer from the position of the constraints where the transition from $\chi_2$ to $\chi_1$ as the dark matter candidate occurs. The former lies above the colorful excluded region as it corresponds to large $\chi_2$ lifetime and, therefore, larger flavon mass since $\tau_{\chi_2}\propto \ms^4$, cf. \cref{eq:dark_3body,eq:semivisible}. Instead, below the excluded region $\tau_{\chi_2}$ is small enough such that it is the lighter dark fermion $\chi_1$ that plays the role of DM.

In the plot, we also show with blue dotted lines expected future improvements of the CMB bounds from the combination of the CMB Stage-4 searches~\cite{CMB-S4:2016ple}, the space mission LiteBIRD~\cite{Matsumura:2013aja}, and the data from the Polarized Radiation Imaging and Spectroscopy Mission (PRISM)~\cite{PRISM:2013fvg} as implemented in Ref.~\cite{Lucca:2019rxf}. The largest improvements, in this case, are expected for the bounds on CMB spectral distortions and they are relevant for the regime of $\tau_{\chi_2}<\tau_{\textrm{recomb.}}$, i.e., the region of the parameter space of the model where $\chi_1$ is the DM candidate observed today. The large $\chi_2$ lifetime regime could, instead, be further constrained in the future by indirect DM searches, e.g. in the Cherenkov Telescope Array (CTA)~\cite{CTAConsortium:2017dvg} or the High-Altitude Water Cherenkov (HAWC) observatory~\cite{Abeysekara:2013tza}, as well as in the proposed neutrino telescopes such as IceCube-Gen2~\cite{IceCube-Gen2:2020qha} or GRAND~\cite{GRAND:2018iaj}, Trinity~\cite{Otte:2019aaf}, and POEMMA~\cite{POEMMA:2020ykm} in the ultra-high energy regime. In the regime of very large $\chi_2$ lifetime, $\tau_{\chi_2}\gtrsim 10^{29}~\textrm{s}$, probing it might remain beyond the reach of currently planned such searches. This region of the parameter space of the model, which is relevant to the upper right corner of the plot, will remain difficult to probe as it predicts a basically stable DM candidate with extremely suppressed interaction rates with the SM species.

We see in the $Q_{\rm FN}(\chi_{2})=6$ and $Q_{\rm FN}(\chi_1)=7$ case that similar areas of viable parameter space have either of the dark-sector particles as the dark matter candidate. The same cannot be said for $Q_{\rm FN}(\chi_{2})=9$ and $Q_{\rm FN}(\chi_1)=10$ shown in the top right panel. Here $\chi_2$ is typically very long-lived and it remains the only DM candidate possible. Note also here that the BBN constraints are absent, this is because the plane does not contain values of $m_\sigma$ and $\Lambda$ for which $\tau_{\chi_2}<0.01\,{\rm s}$. We see in this panel that the decaying dark matter searches from telescopes are the most constraining in this case, where future CMB bounds will not improve upon existing constraints.  

The bottom row of Fig.\ref{fig:multi_ID} demonstrates the effect of having charge differences greater than 1. For both panels $Q_{\rm FN}(\chi_{2})=4$, but $Q_{\rm FN}(\chi_{1})=5$ and $Q_{\rm FN}(\chi_{1})=8$ for the left and right panels respectively. With $\Delta Q_{\rm FN}=1$ on the left, one can see that the experimental probes we consider have no effect on the viable parameters in our setup. We also observe that the BBN constraint is again absent. Here this is because the ${\rm Br\left(\chi_2\rightarrow \chi_1\bar{q}q\right)}$ is negligibly small. From~\eqref{eq:DSSM_flip} it can be estimated that the SM branching ratio becomes relevant only below $10^{10}\,{\rm GeV}$. Moving to the $\Delta Q_{\rm FN}=4$ case on the right, we see a substantially different landscape. Firstly, the lower bound on $m_\sigma$ where the relic abundance is not reachable has moved off the plane. This is simply because $\chi_{1}$ is now much lighter, and the model can now accommodate $m_{\sigma}\rightarrow\chi_1\bar{\chi}_1$ for much smaller $m_\sigma$ values, even when decays with $\chi_2$ in the final state become kinematically forbidden. Furthermore, the higher charge values decrease the lifetime of $\chi_2$ which brings the constraints from telescopes and Lyman-$\alpha$ into more interesting parts of parameter space. Interestingly due to the large visible branching ratio, the limits from CMB and decay products cover larger areas. The same broadening of limits can be observed for the warm dark matter constraints because the lighter $\chi_1$ produced in $\chi_2$ decays is much more boosted. Additionally, ${\rm Br\left(\chi_2\rightarrow \chi_1\bar{q}q\right)}$ is now dominant for much larger values of $\Lambda$ which allows to place constraints based on light element abundances and BBN predictions. 

Last but not least, we comment again on the multi-component case and the relevant cosmological and indirect bounds. As argued above, in the presence of three or more dark fermions $\chi_i$, sequential decays from the heaviest to the lighter ones are expected to play the dominant role in determining the dark sector dynamics. In this case, the interplay between dark and semi-visible decay final states can dominantly affect the decay of only next-to-the-lightest species $\chi_2$. The heavier dark fermions, however, might also be characterized by non-negligible values of semi-visible branching fractions, depending on the choice of the dark FN charges. This could then lead to additional constraints from cosmological probes and indirect detection. Similarly, bounds on the free-streaming length of the DM candidate could be further modified in the presence of sequential chain decays in the multi-component dark sector, as the consecutive such decays could additionally contribute to the DM boost factor. While all these effects could render at least some multi-component DM scenarios strongly constrained, we illustrate above that in the simple two-component DM scenario large regions of the parameter space remain allowed by current bounds while they could partially be probed by future searches.

The sequential decay process can also be affected in the presence of dark fermions carrying the same FN charge, as their mass difference can then be small and make $\chi_i\to \chi_{i-1}\chi_{i-1}\bar{\chi}_{i-1}$ decays kinematically forbidden. In this case, more than a single dark fermion can be produced in the dominant decay mode of $\chi_i$ and some of such lighter fermions can even remain as dark radiation (DR). This opens up an interesting possibility of addressing the cosmological $S_8$ tension in such a model with decaying cold DM (DCDM), see Refs~\cite{Abdalla:2022yfr,DiValentino:2020vvd} for recent reviews. We provide a sample benchmark scenario with three dark fermions $\chi_i$ in \cref{app:massbasis}. In this case, the correct DM relic density is obtained as a sum of the abundances of both massive dark particles $\chi_2$ and $\chi_3$, while $\chi_1$ is DR. The heaviest DM component decays with the lifetime $\tau_{\chi_3}\sim 200~\textrm{Gyr}$ into final states with both the slightly lighter $\chi_2$ fermion and the DR species $\chi_1$. With the mass splitting between the two massive fermions set to $m_{\chi_2}/m_{\chi_3} \simeq 0.992$, such a scenario has been shown to address the cosmological tension without violating other bounds~\cite{Fuss:2022zyt}. As we stress in the appendix, the proposed sample benchmark scenario can be obtained for the flavon mass close to its vev, $m_\sigma\sim v_s$, thus without large fine-tuning of the flavon potential parameters, as well as for the theoretically appealing value of $\Lambda = \Lambda_{\textrm{GUT}} = 10^{15}~\gev$. We leave further detailed studies of the connection between the non-thermal flavon portal to DM and the cosmological $S_8$ tension for future analyses.

\section{Conclusions\label{sec:conclusions}}

In this article we have explored how the solutions to the flavor and dark matter puzzles may be connected. We do this by taking the Froggatt-Nielsen mechanism with a global $U(1)$ symmetry and add new fermions that are uncharged under the Standard Model. This scenario has been investigated previously, but with limited scope. In the literature before us, the focus was on the thermal production of dark matter, requiring fairly low scale ($\sim \tev$) of new symmetry breaking. In this work, we consider a symmetry breaking scale which is so high that the flavon field is never in equilibrium with the Standard Model. Considering, for the first time, the effect that flavon freeze-in and oscillationary production will have on the relic density of dark matter. We show that even in such cases, dark matter relic abundance can be readily achieved with numerous dark FN charge assignments. This can occur at high reheat temperatures $T_R$ and flavon masses $\ms$. Generically, this presents a challenge to experimentally probe this scenario. Despite this, there already exist constraints from Big Bang Nucleosynthesis that limit both the charge of dark matter and $\ms$. For example, we found that for a breaking scale of $\Lambda\sim 10^{15}~\gev$, the largest dark matter charge that can achieve $\OmegaDM=0.12$ would be $Q_{\rm DM}=12$, corresponding to $m_{\rm DM}=0.48~\gev$. 

From a purely model-building perspective, we have argued that, due to the scale separation naturally present in multi-component dark sectors coupled via flavon, production of dark particles will be sequential. Therefore, to good approximation, the results of a relic calculation for one component dark matter will be accurate, even in a setup with many lighter dark fermions. Such multi-component dark sectors may be considered more appealing, given that the Standard Model has three generations. In such cases, a new phenomenological window opens in the form of dark decays. These can occur across large time periods, allowing us to invoke constraints from $\gamma$-ray and neutrino telescopes, Planck data on the Cosmic Microwave Background, Lyman-$\alpha$ measurements and once more, BBN. Where possible, we have shown projections of future probes, and there will be substantial improvements in viable regions of parameter space. There are future experimental searches, that will provide significant improvement to the search for decaying dark matter and that will be relevant for this model, such as CTA and HAWC, as well as future CMB probes and neutrino telescopes. Interestingly, recent studies have pointed to a decaying dark matter scenario with a massive and massless dark sector species in the final state as a potential solution to the cosmological $S_8$ tension. We have illustrated how this can be accommodated in the flavon portal model discussed here, the details of which we leave for future work.

We see this work as a first step in combining the flavour puzzle with the dark matter question for high scale UV completions. With more conventional production mechanisms of dark matter being thoroughly probed by a vast experimental program, appraising the more elusive possibilities is an instructive exercise in determining how we may improve our search strategy. It is encouraging that such models which have such highly suppressed rates of interactions still could produce detectable signals.

This study has been limited by design, in order to keep the discussion and findings clear. We have only focused on the subthermal flavon portal. This allowed us to assess the more experimentally difficult regions to probe, i.e. where $\Lambda$ is high and still come up with interesting insights. An extension of this work would be to analyze the case where the flavon reaches thermal equilibrium at times. We leave this for future work, but note that once the flavon thermalizes, soon the dark sector will thermalize, and one will be back in the highly constrained regime of thermal freeze-out dark matter. Furthermore, we only considered a broken global $U(1)_{\rm FN}$ and assume that the goldstone boson plays no role in the phenomenology. As explained above, this is a legitimate scenario that can be achieved by a number of UV-completions, however, other interesting possibilities exist. For example gauging the $U(1)_{\rm FN}$ symmetry would introduce a new vector boson and interactions between the SM and the dark sector. Typically one would expect this to change the phenomenology quite drastically, this is worth investigating in future, especially since the expected cosmic string network formed from symmetry breaking provides an upper limit on $\Lambda$.\\

This work shows that the mass hierarchy in the SM and the origin of dark matter may stem from the same extension of the SM. It is encouraging to note that even with high breaking scales of the flavor symmetry $\sim\Lambda_{\rm GUT}$, the model that we consider can correctly reproduce the observed DM relic abundance. While the predicted DM particles are generically extremely weakly coupled to the SM and easily avoid (in)direct bounds, the scenario is not without constraints from phenomenological probes. This is true when one allows for the possibility that there are more than one generation in the dark sector with the relative hierarchy driven by varying coupling strengths to the flavon fields, therefore mimicking a corresponding mechanism proposed for the SM. The non-thermal flavon portal to DM could similarly be applied to more complicated scenarios going beyond the simplest $U(1)_{\textrm{FN}}$ horizontal symmetry.

\section*{Acknowledgements} 
AC would like to thank Jan Heisig for discussions about indirect probes of decaying dark matter. ST would like to thank Benjamin Lillard, Michael Ratz, and Tim M. P. Tait for useful discussions about non-thermal flavon production mechanisms. This work is supported by the grant ``AstroCeNT: Particle Astrophysics Science and Technology Centre" carried out within the International Research Agendas programme of the Foundation for Polish Science financed by the European Union under the European Regional Development Fund. LR is also supported in part by the National Science Centre, Poland, research grant No. 2015/18/A/ST2/00748. ST is also supported in part by the National Science Centre, Poland, research grant No. 2021/42/E/ST2/00031. ST is supported in part by the Polish Ministry of Science and Higher Education through its scholarship for young and outstanding scientists (decision no 1190/E-78/STYP/14/2019). 
ST acknowledges the support of the Institut Pascal at Université Paris-Saclay during the Paris-Saclay Astroparticle Symposium 2022, with the support of the P2IO Laboratory of Excellence (program “Investissements d’avenir” ANR-11-IDEX-0003-01 Paris-Saclay and ANR-10-LABX-0038), the P2I axis of the Graduate School Physics of Université Paris-Saclay, as well as IJCLab, CEA, IPhT, APPEC, the IN2P3 master projet UCMN and ANR-11-IDEX-0003-01 Paris-Saclay and ANR-10-LABX-0038.

\appendix 
\section{Dark fermion masses in multi-component scenarios\label{app:massbasis}}

The rotation to the mass basis in the dark fermion sector introduces additional corrections to the corresponding flavon couplings that we discuss below for a specific case of two-component DM scenario. We also present a simple benchmark scenario with two-component DM and additional dark radiation species which could address the cosmological $S_8$ tension.

\subsection{Two-component case} 

In the two-component DM scenario, in which $\chi_1$ and $\chi_2$ are dark flavor states with $Q_{\chi_1} = Q_{\chi_2} + n$, the diagonalization of the mass matrix obtained from \cref{eq:Lchi1} for $y<1$ and $n\geq 1$ leads to 
\begin{equation}
m_{1} \simeq m_{11}\,(1-y^2),\hspace{1cm} m_{2} \simeq m_{22}\,(1+y^2\epsilon^{2n}),
\end{equation}
where $m_{ii}$ are given by \cref{eq:mdm_eq} and we assumed $y^2\,\epsilon^{2n}\ll 1$. As can be seen, the masses of the relevant species can be well approximated with the diagonal terms in \cref{eq:Lchi1}, unless $y\simeq 1$. In the latter case, the lighter of the dark species becomes dark radiation. 

In the limit of $y\,\epsilon^n\ll 1$, the mixing angle between the dark fermions reads $\sin\theta \simeq y\,\epsilon^n$. The corresponding corrections to the flavon couplings in \cref{eq:flavon_chichi_int} are approximately given by
\begin{eqnarray}
g^\chi_{11}\to g^\chi_{11} & \times & \frac{n_{11}}{n_{11}+1}\,\left(1+2y - \frac{2y}{n_{12}} + y^2 - \frac{2y^2}{n_{11}}\right)\simeq g^\chi_{11}\,(1+y)^2,\\
g^\chi_{12}\to g^\chi_{12} & \times & \frac{n_{12}}{n_{12}+1}\,\left(1+y - \frac{y}{n_{12}} + \mathcal{O}(\epsilon^{2n})\right)\simeq g^\chi_{12}\,(1+y),\\
g^\chi_{22}\to g^\chi_{22} & \times & \frac{n_{22}}{n_{22}+1}\,\left(1- 2\,y\,\epsilon^{2n} - \frac{2\,y\,\epsilon^{2n}}{n_{22}} + \mathcal{O}(\epsilon^{4n})\right)\simeq g^\chi_{22} \ .
\end{eqnarray}
Importantly, the interaction rate of the heavier dark fermion, $g_{22}^\chi$, is only very mildly affected by the mixing and, therefore, the dominant contribution to the DM relic density produced out of late-time flavon decay, $\sigma\to \bar{\chi}_2\chi_2$, is not sensitive much to the details of the mixing and the particular choice of the parameter $y$. The impact of mixing is more substantial on the $\chi_2$ decay width for the process, $\chi_2 \to \chi_1\bar{\chi_1}\chi_1$, which is sensitive to both $(g_{12}^\chi)^2$ and $(g_{11}^\chi)^2$ and receives a correction of order $(1+y)^6$. For instance, for $y=1/3$, as assumed in our plots, the correction due to the rotation to the mass basis results in about a factor of $6$ smaller $\chi_2$ lifetime which we take into account in our analysis. 
 
\subsection{Two-component dark matter with dark radiation}

In this section, we briefly present a benchmark scenario with three dark fermions $\chi_i$ out of which two are massive with an almost degenerate mass, $m_{\chi_3}\simeq m_{\chi_2}$, and one turns out to be massless dark radiation with vanishing $m_{\chi_1}$. This can be connected to the decaying cold DM solution to the $S_8$ cosmological tension, as mentioned in \cref{sec:multicomponentresults}.

Let us assume for simplicity that two out of the three dark fermions have the same dark FN charge which is much smaller than the charge of the remaining fermion, $Q_{\chi_1}\gg Q_{\chi_2} = Q_{\chi_3}$. This implies that the flavon decay branching fraction into $\chi_1$ will be highly suppressed. We also need to specify the order-one coefficients $y_{ij}$, cf. \cref{eq:Lchi1}. For concreteness, we take $y_{ii} = 1$, $y_{23} = y_{32} = 1$, and $y_{12} = -y_{21} = y_{13} = -y_{31} = y \neq 1$. One can then check that the mass matrix for the dark fermions diagonalizes with vanishing $m_{1}$ and 
\begin{equation}
m_{2,3} = \epsilon^{2\,Q_{\chi_3}}\,\Lambda\,\left[1 + \frac{\epsilon^{2\,Q_{\chi_1}}}{2} \pm \sqrt{1 + 2y^2 - \epsilon^{2\,Q_{\chi_1}} + \frac{e^{4\,Q_{\chi_1}}}{4}}\right] \simeq \epsilon^{2\,Q_{\chi_3}}\,\Lambda\,\left[1 \pm \sqrt{1 + 2y^2}\right]\ .
\end{equation}
The lightest dark fermion state $\chi_1$, therefore, plays the role of dark radiation whose production rate in flavon decays in the early Universe can be made negligible for large $Q_{\chi_1}$ charges. Further assuming that $y\simeq i[1/\sqrt{2} -c]$ with $c\ll 1$, one obtains $m_{2}\simeq m_3$, i.e., both heavier states are almost mass degenerate. In particular, the mass ratio $m_{\chi_2}/m_{\chi_3}\simeq 0.992$ can be obtained for $c\simeq 5.7\times 10^{-6}$.

Since both heavier states have the same dark FN charge, they will be produced in the early Universe with a very similar abundance such that $\Omega_{\chi_2}\simeq \Omega_{\chi_3}\simeq \Omega_{\textrm{DM}}/2$. They can both decay into either SM particles or lighter dark states. Given a very small mass splitting between these dark fermions, the available dark decay channels for the heaviest species are $\chi_3\to \chi_2 \chi_1\bar{\chi}_1$, $\chi_3\to \chi_1 \chi_1\bar{\chi}_2$, and $\chi_3\to \chi_1\chi_1\bar{\chi}_1$. We note that the branching fraction for the last process is suppressed by an additional factor of $\epsilon^{Q_{\chi_1}-Q_{\chi_3}}\ll 1$. The dark fermion $\chi_3$ can also decay semi-visibly, dominantly into $\chi_3 \to \chi_2q\bar{q}$ final states, although the fully invisible dark decays into $\chi_2$ and $\chi_1$ dominate for large values of $\Lambda$, cf. \cref{eq:DSSM_flip}. In particular, for $Q_{\chi_3} = 5$, $Q_{\chi_1} = 8$, and $\Lambda\sim \Lambda_{\textrm{GUT}}\sim 10^{15}~\gev$, such semi-visible decays are highly suppressed for $\chi_3$ which, in turn, will decay invisibly with $\tau_{\chi_3}\sim 200~\textrm{Gyr}$ for the flavon mass $m_\sigma \sim 10^{14}~\gev$. We note that $m_\sigma\sim v_s = \epsilon\Lambda$ as one would naturally expect without fine-tuning the parameters of the flavon potential. In this case, the mass of both massive dark fermions is of order $m_{\chi_3}\simeq m_{\chi_2}\sim 10^{8}~\gev$ and their combined relic density from out-of-equilibrium flavon decays matches the observed DM relic density for $T_R\sim 10^{12}~\gev$. 

As can be seen, the three-component non-thermal flavon portal to heavy DM can reproduce the benchmark scenario identified to address the cosmological $S_8$ tension with $\chi_2$ playing the role of a massive lighter partner of a decaying $\chi_3$ DM component and $\chi_1$ being dark radiation. Notably, both $\chi_3$ and $\chi_2$ will constitute DM at the present time with roughly equal abundances. The lighter dark fermion $\chi_2$ will be effectively stable with a lifetime much exceeding the age of the Universe so only half of the DM species will decay ($\chi_3$). This should be taken into account in the precise modeling of this scenario and its impact on $S_8$. While we present here a simple possible benchmark scenario to realize this phenomenology, as a proof of principle, we also stress that in a more thorough treatment one could search for scenarios with semi-visible branching fraction of $\chi_3$ being not completely suppressed. This could introduce additional and independent signatures of such models in future DM ID searches. We leave this for future studies. 

\section{Different dark charges between left- and right-handed dark fermions\label{app:darkcharges}}

Throughout the article, we have followed a simplifying assumption that left- and right-handed dark fermions carry the same charge under the $U(1)_{\textrm{FN}}$ group, $Q_{\chi_i,L} = Q_{\chi_i,R}$. We note that essential aspects of the DM phenomenology discussed in our study depend on the combination of the dark charges $n_{ij}^\chi$, given in \cref{eq:chargeschi}, rather than on individual values of $Q_{\chi_i}$. As a result, many features of the BSM scenario that we present are still expected to hold in a more general case, where $Q_{\chi_i,L} \neq Q_{\chi_i,R}$, provided that the sum of the dark charges reproduces the same value of $n_{ij}^\chi$. However, going beyond the left-right symmetric regime might also lead to interesting discrepancies from this simple picture. While we leave a full discussion of this issue for the future, we highlight below a few such modifications and further directions to explore that could arise in the discussion of our BSM scenario for more general charge assignments in the dark sector.

As a first example, in \cref{eq:Omegaheuristic} we have made a heuristic argument about the maximum possible relic density of the lighter of the two dark fermions, $\chi_1$, in the two-component DM case. We argue there that for a unit charge difference between the two, $\Delta Q_\chi = 1$, one expects no more than about a $3\%$ contribution from $\chi_1$ to the total DM relic abundance dominated by the heavier species $\chi_2$. This motivates why a one-component dark sector is a good approximation to describe the phenomenology of the multi-component case. We note, however, that if the dark Yukawa matrix or the charge configuration was different, the above relation would be slightly different. For example, if $\mathcal{Q}_{\chi_{i,L}}\neq\mathcal{Q}_{\chi_{i,R}}=0$, the ratio could be as high as $\Omega_{\chi_1}\approx 0.082\,\Omega_{\chi_2}$. Even in this case, however, the relic density contribution from the lighter dark species would remain less than $10\%$, with the heaviest dark fermion still dominating in the total DM abundance. Instead, similar relic abundances for both the dark fermions can be obtained when they share the same FN charge, as we illustrate in \cref{app:massbasis}.

More substantial differences with respect to the left-right symmetric scenario can arise when considering dark fermion decays. If the dark charges differ significantly between the chiral states, one of the left- or right-handed fermion has larger interaction rates with the flavon field and with other dark fermions. Let us assume for concreteness that $Q_{\chi_i,R} \gg Q_{\chi_i,L}$ for a given dark fermion $\chi_i$ and that both charges are equal for some other dark fermion $\chi_j$, i.e. $Q_{\chi_j,L} = Q_{\chi_j,R}$. We will also assume that $Q_{\chi_i,L} = Q_{\chi_j,L}$ for simplicity. The last assumption leads to a clear mass hierarchy between the dark fermions $m_{\chi_j} \gg m_{\chi_i}$, given that their masses in \cref{eq:mdm_eq} depend on the sum of the dark charges that determine the $n_{ii}^\chi$ and $n_{jj}^\chi$ parameters. In particular, the mass of $\chi_i$ is mainly set by the larger of the charges, $Q_{\chi_i,R}$. The heavier of the two dark fermions can decay into the lighter one, $\chi_j\to\chi_i\,\chi_i\,\bar{\chi}_i$ or $\chi_j\to\chi_i\,q\,\bar{q}$. The relevant decay amplitude averaged over initial spins is mainly driven by the combination of charges for left-handed $\chi_i$ and right-handed $\chi_j$, i.e., $Q_{\chi_i,L}+Q_{\chi_j,R}$, while the other sum is much larger and leads to a suppressed contribution. As a result, the decay rate of $\chi_j$ into $\chi_i$ is largely independent of $Q_{\chi_i,R}$ and, therefore, the $\chi_i$ mass. This is similarly true for the flavon decay, $\sigma \to \chi_i\bar{\chi}_j$.

This can affect the sequential nature of both the dark fermion production process from flavon decays and the decay processes in the dark fermion sector. These are characteristic for the left-right symmetric scenario. Beyond this regime, both the flavon and heavy-fermion decays can lead to the lighter DM species $\chi_i$ with a substantial relic density even in the presence of much less abundant intermediate heavier dark fermions. In particular, if the initial abundance of the heavier dark fermion is efficiently transformed into the DM relic density of the much lighter one, the maximum allowed reheating temperature $T_R$, which determines the flavon yield, can also grow larger. In addition, reducing the lifetime of the heavier dark fermion $\chi_j$ could weaken otherwise increasingly more stringent astrophysical and cosmological bounds obtained for the growing charge difference between dark fermions, cf. \cref{fig:multi_ID} (bottom right panel) for the symmetric case. Phenomenologically viable scenarios can then appear more easily that are characterized by a large mass hierarchy in the dark sector, similar to the SM, and by a natural value of the flavon mass, $m_\sigma\sim v_S$, at high energy scales $\Lambda\gg \textrm{TeV}$.

\bibliography{flavon_BBN}

\end{document}